\documentclass[twocolumn]{raa}           
\usepackage{graphicx,times}  
\usepackage{natbib}
\usepackage{amssymb,amsmath}
\usepackage{xcolor}
\bibpunct{(}{)}{;}{a}{}{,}
\usepackage{threeparttable}
\usepackage{textcomp}
\usepackage[a4paper=true, pagebackref=true]{hyperref}
\hypersetup{pdftitle = The title of my PDF, pdfauthor = My name, pdfsubject= The subject, pdfkeywords = keyword1 keyword2 keyword3} 
\hypersetup{colorlinks = true, linkcolor = green, anchorcolor = red, citecolor = blue, filecolor = red, pagecolor = red, urlcolor = red}

\begin{document}
\title{The Column Density Structure of Orion A Depicted by N-PDF}
\volnopage{ {\bf 20XX} Vol.\ {\bf X} No. {\bf XX}, 000--000}
\setcounter{page}{1}
\author{Yuehui Ma \inst{1, 2} \and Hongchi Wang \inst{1} \and Chong Li \inst{1, 2} \and Ji Yang \inst{1}} 
\institute{Purple Mountain Observatory and Key Laboratory of Radio Astronomy, Chinese Academy of Sciences,
10 Yuanhua Road, Nanjing 210033; {\it mayh@pmo.ac.cn}\\
	\and 
	    University of Chinese Academy of Sciences, 19A Yuquan Road, Shijingshan District, Beijing 100049, China\\}
\vs \no
{\small Received 2019/10/22; accepted 2019/12/10}

\abstract{We have conducted a large-field simultaneous survey of $^{12}$CO, $^{13}$CO, and C$^{18}$O $J=1-0$ emission toward the Orion A giant molecular cloud (GMC) with a sky coverage of $\sim$ 4.4 deg$^2$ using the PMO-13.7 m millimeter-wavelength telescope. We use  the probability distribution function of the column density (N-PDF) to investigate the distribution of molecular hydrogen in the Orion A GMC. The H$_2$ column density, derived from the $^{13}$CO emission, of the GMC is dominated by log-normal distribution in the range from $\sim$4$\times10^{21}$ to $\sim$1.5$\times10^{23}$ cm$^{-2}$ with excesses both at the low-density and high-density ends. The excess of the low-density end is possibly caused by an extended and low-temperature ($\sim$10 K) component with velocities in the range of 5$-$8 km s$^{-1}$. Compared with the northern sub-regions, the southern sub-regions of the Orion A GMC contain less gas with column density in $N_{H_2} > 1.25\times 10^{22}\ \rm{cm}^{-2}$. The dispersions of the N-PDFs of the sub-regions are found to correlate with the evolutionary stages of the clouds across the Orion A GMC. The structure hierarchy of Orion A GMC is explored with the DENDROGRAM algorithm, and it is found that the GMC is composed of two branches. All structures except one in the tree have virial parameters less than 2, indicating self-gravity is important on the spatial scales from $\sim$0.3 to $\sim$4 pc. Although power-laws and departures from log-normal distributions are found at the high-density end of N-PDFs of active star-forming regions, the N-PDFs of structures in the Orion A GMC are predominantly log-normal on scales from R$\sim$0.4 to 4 pc.
}

\authorrunning{Y.-H. Ma et al.}            
\titlerunning{N-PDFs of Structures of Orion A GMC}  
\maketitle
 
\section{Introduction}         
\label{sec:1}
Stars form in the cold and dense molecular clouds. The cloud structure is affected by various physical processes, such as turbulence, cloud-cloud collisions, and feedbacks from massive stars. The probability distribution function (PDF) is a simple but potent tool to depict the structure of molecular clouds, therefore, providing an efffective means to investigate the various physical processes that influence the structures of molecular clouds. Theoretical studies \citep{Vazquez-Semadeni1994, Padoan1997, Klessen2000} of supersonic turbulence suggest that the volume density PDFs ($\rho$-PDFs) of molecular clouds exhibit log-normal shapes when self-gravity is not important. This characteristic has been attributed to the influence of some steady and multiple independent dynamical events that shape the molecular clouds \citep{Mckee2007}. When star formation activities occur, the $\rho$-PDF is strongly affected by gravity in addition to turbulence, resulting in excess above the log-normal distribution at the high density end \citep{Kainulainen2014, Federrath2013}. $\rho$-PDF is also proved to be a reliable indicator for the star formation rate and efficiency of molecular clouds in different star formation models \citep{Elmegreen2008, Padoan2014}. However, it is challenging to constrain the underlying $\rho$-PDFs of molecular clouds with observations because of the projection effect along the line-of-sight. Instead, a lot of theoretical and observational studies focused on the column density PDFs (N-PDFs). Simulations have suggested an evolutionary trend of molecular clouds from the turbulent dominated stages to the gravity dominated star-forming stages \citep{Ballesteros-Paredes2011, Kritsuk2011}. The gravity first broadens the log-normal N-PDF at the beginning and then produces a power-law tail in the N-PDF \citep{Ballesteros-Paredes2011}. The slope of the power-law tail develops over time and gets shallower as the star formation efficiency (SFE) increases.

Observationally, three tracers of column density are widely used, which are the near infrared dust extinction, the far-infrared dust emission, and the optically thin molecular line emission. Comparatively, dust extinction probes the largest dynamic range of column density and less relies on models \citep{Goodman2009a}. However, the gas tracer can provide us the kinematic information of the molecular clouds, which helps to disentangle the structure of molecular clouds. Log-normal shaped N-PDFs derived with the three tracers have been observed in the Perseus cloud \citep{Goodman2009a}, and power-law N-PDFs have been observed in active star-forming regions \citep{Kainulainen2009, Schneider2013, Lombardi2015}. In the observations mentioned above, quiescent clouds often exhibit log-normal N-PDFs in the whole dynamic range or only have modest excesses at the high density ends. In contrast, clouds active in star formation have N-PDFs with prominent non-log-normal or power-law high-density tails arising from the dense star-forming regions.  

As the nearest massive star-forming region (d = 414 pc, \citealt{Menten2007}), the Orion A GMC is notable for its complex hierarchical filamentary structure and the intense star formation activities \citep{Bally2008}. There have been plenty of large-scale surveys toward the Orion A GMC at multi-wavelengths \citep{Bally1987, Sakamoto1994, Johnstone1999, Shimajiri2011, Megeath2012, Fischer2013, Ripple2013, Stutz2013, Berne2014, Shimajiri2014, Nishimura2015, Stutz2015, Groschedl2018, Hacar2018, Kong2018, Suri2019}.  The GMC is mainly composed of a large and dense integral-shaped-filament (ISF, \citealt{Bally1987}) in the northern part that contains star-forming regions OMC 1-4, and a relatively extended and less dense tail in the southern part that covers the LDN 1641 region. \cite{Kong2018} have made a brief review of the surveys that covers the ISF in their table 1. The broad dynamic range of column densities revealed by the previous studies makes the GMC an ideal case for N-PDF analysis. The N-PDF of Orion A GMC has been derived with multi-wavelength observations \citep{Kainulainen2009, Lombardi2015, Stutz2015, Berne2014}, most of them presenting a N-PDF of power-law shape. However, most of the previous N-PDF studies of Orion A are limited to the ISF region except for the $Herschel$ observations \citep{Stutz2015} (S15 hereafter). S15 found that all the N-PDFs in different subregions of the Orion A GMC exhibit power-law distributions, and the power-law exponent of the N-PDF correlates with the fraction of the Class 0 protostars. For a panoramic view of the Orion A GMC, it is valuable to study the N-PDF of the entire Orion A GMC using molecular line emission and make comparison with N-PDFs derived using dust emission. Furthermore, for the investigation of the relation between N-PDF forms and the underlying physics, it is necessary to find out the dependence of the N-PDF forms and the role of self-gravity on spatial scales. There are several automatic algorithms that can identify structures in molecular clouds, like CLUMPFIND \citep{Williams1994}, GAUSSCLUMPS \citep{Stutzki1990}, FELLWALKER \citep{Berry2015}, CPROPS \citep{Rosolowsky2006}, and DENDROGRAM \citep{Rosolowsky2008}. Among the algorithms, DENDROGRAM can reveal the hierarchical nature of molecular clouds, and it is relatively insensitive to the input parameters.   

In this work, we present a detailed analysis of N-PDF of the Orion A GMC using the $^{12}$CO and $^{13}$CO $J=1-0$ emission line data from a large-scale ($\sim$ 4.4 deg$^2$) survey toward the GMC. The N-PDFs of the hierarchical structures and the role of gravity on different spatial scales in the Orion A GMC have been studied using the DENDROGRAM technique. In Section \ref{sec:2}, we describe the observations and the data reduction processes. The main results are presented in Section \ref{sec:3}. The N-PDF properties and the importance of gravity in the GMC on different spatial scales are discussed in Section \ref{sec:4}. We summarize our results and make conclusions in Section \ref{sec:5}.

\section{Observations and data reduction} \label{sec:2}
\subsection{Observations}
The observations were made using the PMO-13.7 m millimeter-wavelength telescope located at Delingha in China during 2011 June as a pilot survey of the Milky Way Imaging Scroll Painting project \footnote{\url{http://www.radioast.nsdc.cn/mwisp.php}} \citep{Su2019}. The PMO-13.7m telescope is equipped with a nine-beam Superconducting Spectroscopic Array Receiver (SSAR) that works in the 85-115 GHz frequency band \citep{Shan2012}. With a carefully selected local oscillator (LO) frequency, the $^{12}$CO, $^{13}$CO, and C$^{18}$O $J=1-0$ line emission data can be obtained simultaneously with this instrument. The upper sideband contains the $^{12}$CO $J=1-0$ and the lower sideband contains the $^{13}$CO and C$^{18}$O $J=1-0$ line emission. For each sideband, a Superconductor-Insulator-Superconductor (SIS) mixer works as the front end of the receiver. The back-end of the receiver is a Fast Fourier Transform Spectrometer (FFTS) with 16,384 frequency channels and a total bandwidth of 1 GHz, providing a velocity resolution of 0.17 km s$^{-1}$ at 110 GHz. At 110 GHz and 115 GHz, the half-power beam width (HPBW) of the telescope is about 52 $\arcsec$ and 50$\arcsec$, respectively. 

\begin{figure}
	\centering
	\includegraphics[trim = 0cm 2.5cm 0cm 4cm, width=0.55\linewidth, clip]{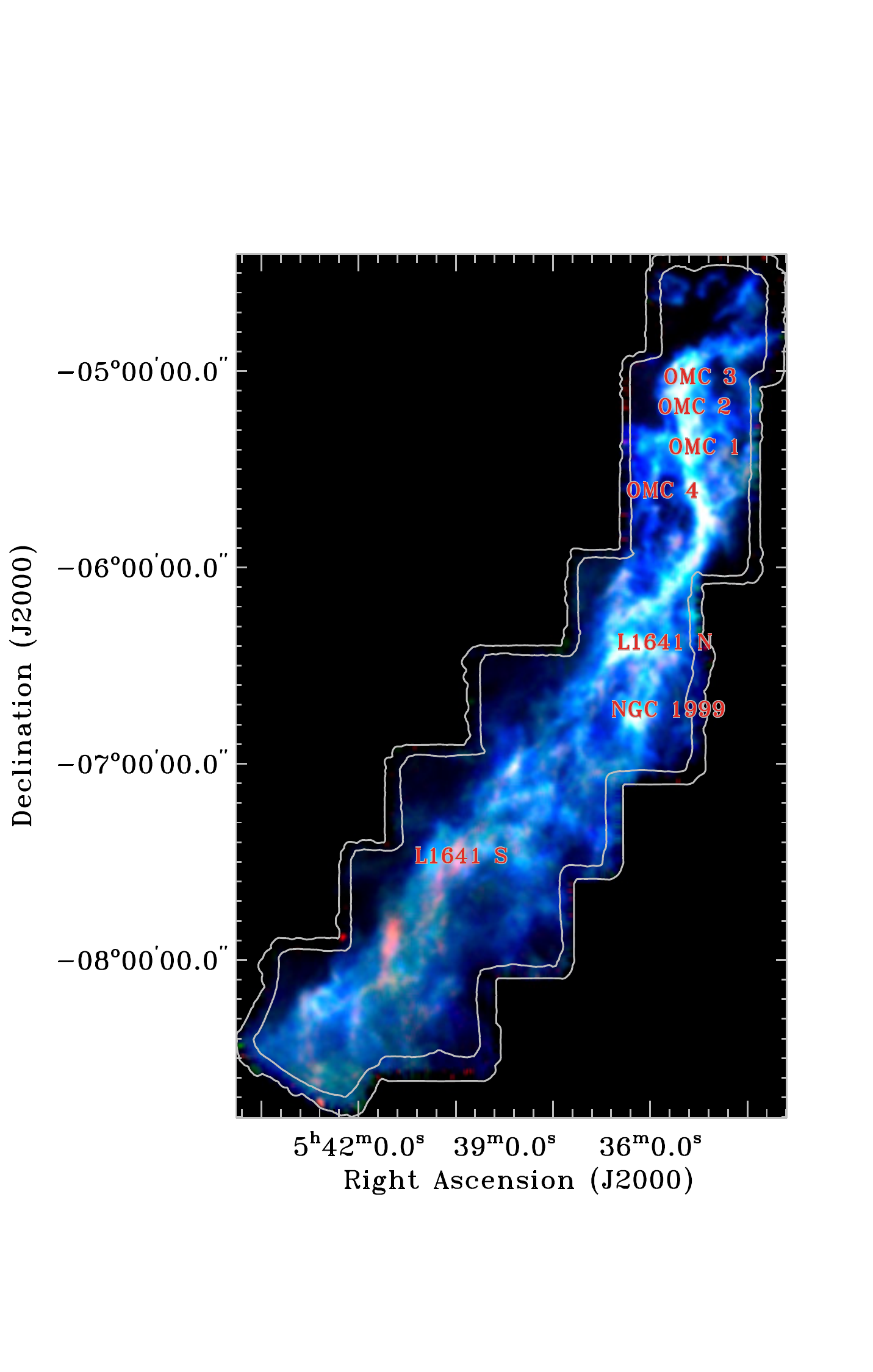}
	\caption{Color-coded image of integrated intensity of the Orion A GMC, with the $^{12}$CO emission in the velocity range from $-$10 to 20 km s$^{-1}$ in blue, the $^{13}$CO emission from 0 to 18 km s$^{-1}$ in green, and the C$^{18}$O emission from 0 to 14 km s$^{-1}$ in red. The grey lines outline the noisy edge of the surveyed area that has been trimmed in the analysis. Active star-forming regions in the GMC are indicated with red letters.}
	\label{Fig1}
\end{figure}

The observations were conducted in the position-switch on-the-fly (OTF) mode. The Orion A GMC is divided into twelve 30$\arcmin\times$30$\arcmin$ cells. For each cell two scannings were made along the directions of right ascension and declination, respectively. The scanning rate of the observation is 50$\arcsec$ per second and the dump time is 0.3 s. The antenna temperature is calibrated according to $\rm{T_{MB}=T^{*}_{A}/\eta_{MB}}$ during the data reduction processes. According to the status report of the PMO-13.7 m telescope, the main beam efficiencies are 0.44 and 0.48 at the $^{\rm 12}$CO $J=1-0$ and the $^{\rm 13}$CO $J=1-0$ wavelengths, respectively. We use the GILDAS/CLASS software package to reduce the data, which includes the subtraction of a second order baseline from each spectrum and the re-griding of the raw data. At the positions where multiple spectra are taken at different times, a combination of the spectra is made using the reciprocal square of the RMS noise level of each spectrum as the weight. The final reduced data cube covers a sky area of $\sim$4.4 deg$^2$, which contains the majority CO emission of the GMC as revealed by \cite{Bally1987}. However, the spectra at the edge of the surveyed area have low signal-to-noise ratios and they are removed in the analysis of this work. The trimmed edge area, indicated with grey lines in Figure \ref{Fig1}, accounts for about twenty percent of the scanned area. The final median RMS noise level of the $\sim$3.5 deg$^2$ effective region is 0.61 K per channel at 115 GHz and 0.37 K per channel at 110 GHz.

\subsection{Method of Structure Identification} \label{sec:4.1}
In this work, we use the DENDROGRAM algorithm to extract ``tree" structures from the H$_2$ column density map and investigate the properties of the N-PDFs of the hierarchical structures and the importance of self-gravity at various spatial scales in Orion A in Section \ref{sec:4}. The threshold and the difference between two separate column density peaks for the identification of ``leaf" of the tree structure are set to be 2.0 and 3.0 times the median noise level of N$_{H_2}$ (see text in Section \ref{sec:3.1}), respectively. The minimum pixel number of a ``leaf" structure is set to 200, which physically corresponds to a clump of the size of $\sim$0.8$\times$0.8 pc$^2$ at the cloud distance. The outputs of the DENDROGRAM provide masks of the identified structures.

We evaluate the importance of self-gravity in a molecular structure using the virial parameter $\alpha_{vir}$, which measures the ratio between the total kinetic energy and the gravitational energy. The widely used definition of $\alpha_{vir}$ is $5\sigma_v^2R/GM_c$, where $\sigma_v$ is the one-dimensional velocity dispersion, G is the gravitational constant, $R$ and $M$ are the effective radius and mass of the structure, respectively. The critical value of $\alpha_{vir}$ for a marginally bound structure is 2 \citep{Kauffmann2013}. The angular size of a structure is defined as the geometric mean of the FWHMs of its major and minor axes which can be obtained in the output table of the DENDROGRAM algorithm, and the LTE mass can be obtained using the total column density within the periphery of the structure. The velocity dispersion of a molecular structure is obtained by fitting a gaussian function to the average spectrum of the $^{13}$CO emission. 

\section{Results}\label{sec:3}
Figure \ref{Fig1} presents the composite intensity map of the $^{12}$CO, $^{13}$CO, and C$^{18}$O $J=1-0$ emission of the Orion A GMC. The $^{12}$CO and $^{13}$CO emission in the region are much more extended than the C$^{18}$O emission, which is mostly concentrated in the active star-forming regions indicated in Figure \ref{Fig1}. Compared with the emission of the $^{12}$CO and $^{13}$CO, the C$^{18}$O emission traces positions with high column density, therefore, only providing a limited dynamical range for the N-PDF study. In the present work, we use the $^{12}$CO and $^{13}$CO data for the N-PDF analysis and the analysis of the C$^{18}$O data is deferred to a future paper.   

\subsection{Calculation of the Column Density}
\label{sec:3.1}
The $^{12}$CO emission is usually optically thick in typical molecular cloud environments, while the $^{13}$CO emission is often optically thin. Providing the molecular clouds are under the local thermodynamic equilibrium (LTE) conditions and the molecules are uniformly excited, the column density of the clouds can be derived with the $^{12}$CO and $^{13}$CO spectra. 

Firstly, the excitation temperature can be obtained with the peak brightness temperature of the optically thick $^{12}$CO line \citep{Li2018},
\begin{equation}
	T_{ex} = 5.532[ln(1+\dfrac{5.532}{T_{peak}+0.819})]^{-1}
	\label{eq:1}
\end{equation}
where $T_{peak}$ is the peak brightness temperature of the $^{\rm 12}$CO $J=1-0$ line. The optical depth and the column density of the $^{13}$CO $J=1-0$ emission can be derived through \citep{Pineda2010, Li2018}
\begin{equation}
\tau^{13}_v = -ln \left\{1 - \dfrac{T_{MB}(^{13}CO)}{5.29[J(T_{ex})-0.164]}\right\}
\label{eq:2}
\end{equation}

\begin{equation}
N_{^{13}CO}=2.42\times10^{14}\frac{T_{ex}+0.88}{1-e^{-5.29/T_{ex}}}\int{\tau^{13}_v dv}.
\label{eq:3}
\end{equation}
where $T_{ex}$ is the excitation temperature, and $J(T_{ex}) = [exp(5.29/T_{ex})-1]^{-1}$. According to \cite{Pineda2010}, when the optical depth of the $^{13}$CO emission is small, $\tau_{v}^{13} < 1$, the integral item $T_{ex}\int{\tau}dv$ can be approximated to $\tau_0/(1-e^{-\tau_0 }) \int T_{mb}\ dv$, where $\tau_0$  is the peak optical depth of the $^{13}$CO emission line. Providing that the abundance ratios [$^{12}$C]$/$[$^{13}$C] $=$ 69 \citep{Wilson1999} and [CO]$/$[H] $\sim$ 10$^{-4}$ \citep{Solomon1972, Herbst1973}, the [$H_{2}/^{13}$CO] ratio is expected to be $7\times10^{5}$, the column density of molecular hydrogen can be obtained with the following formula \citep{Li2018}
\begin{equation}
N_{H_2} = 1.694\times10^{20}\frac{\tau_0}{1-e^{-\tau_0}}\frac{1+0.88/T_{ex}}{1-e^{-5.29/T_{ex}}}\int{T_{MB}(^{13}CO) dv},
\label{eq:4}
\end{equation}

The peak brightness temperature of $^{12}$CO emission and the integrated intensity of $^{13}$CO emission are calculated in the velocity range from 0 to 18 km s$^{-1}$. Observationally, molecular clouds are defined as contiguous structures above a fixed brightness temperature in the position-position-velocity space. The molecular line emission from the clouds should have a certain width of velocity dispersion ($\sigma_v\sim 1$ km s$^{-1}$). To avoid missing weak signals and to ensure an adequate signal-to-noise ratio, we only use the spectra that have at least five contiguous velocity channels with intensities higher than 3 $\sigma_{rms}$ for $^{12}CO$ for the calculation of excitation temperature, and 1.5$\sigma_{rms}$ for $^{13}CO$ for the calculation of column density. According to the above criteria, the signal-to-noise ratio of selected $^{13}$CO spectra is above 3.35, which is sufficiently high to exclude the noisy channels. The three-sigma detection limit of the H$_2$ column density is $\sim$1.5$\times$10$^{21}$ cm$^{-2}$ under the criteria. 

\subsection{Basic Properties of the Orion A GMC}
\begin{figure} [!htb]
	\centering
	\begin{minipage}[t]{0.45\linewidth}
		\centering
		\includegraphics[trim = 0cm 5cm 1cm 4.5cm, width=\linewidth, clip]{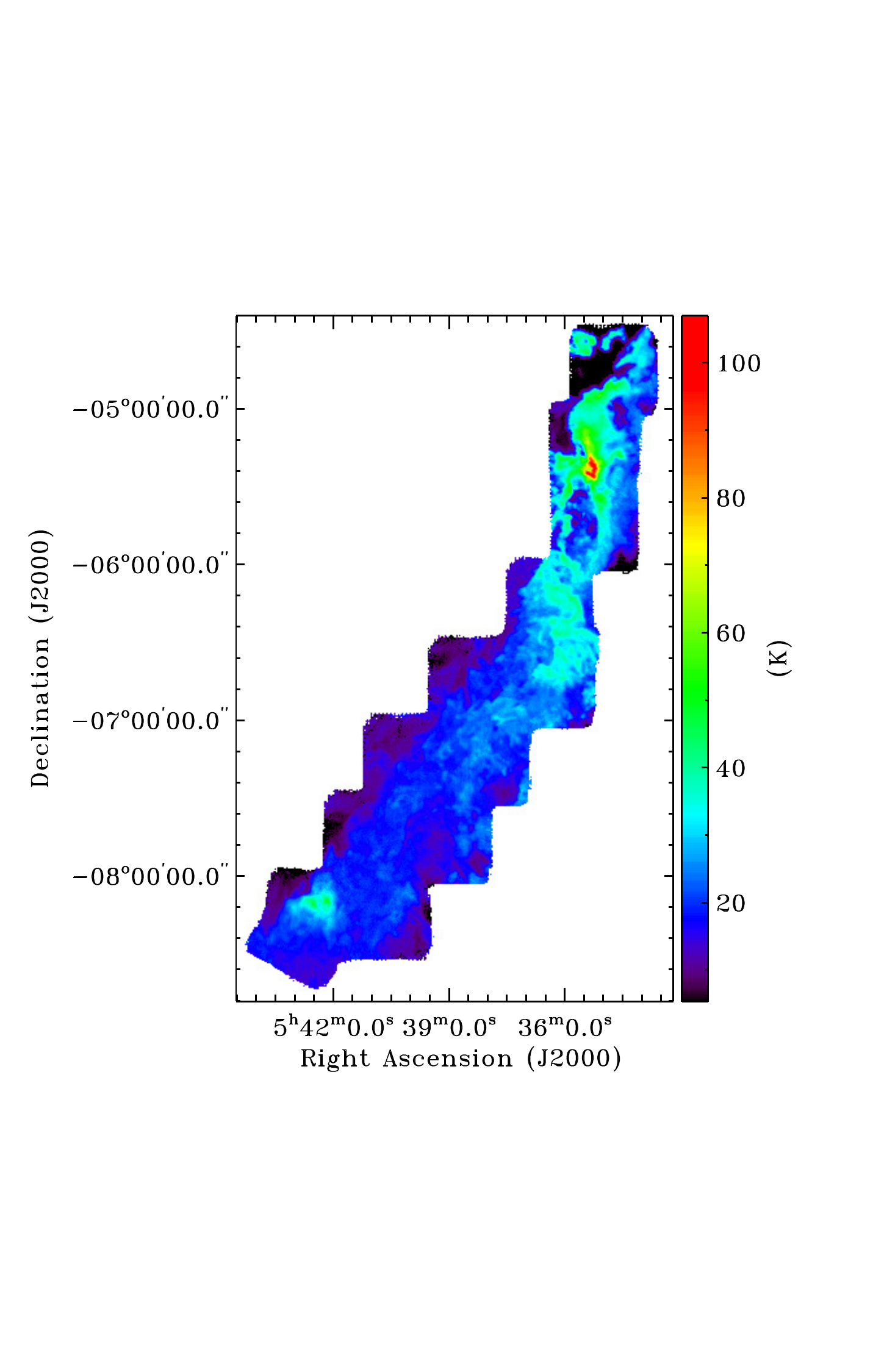}
		\\(a)
	\end{minipage}
	\begin{minipage}[t]{0.45\textwidth}
		\centering
		\includegraphics[trim = 0cm 5cm 1cm 4.5cm, width=\linewidth, clip]{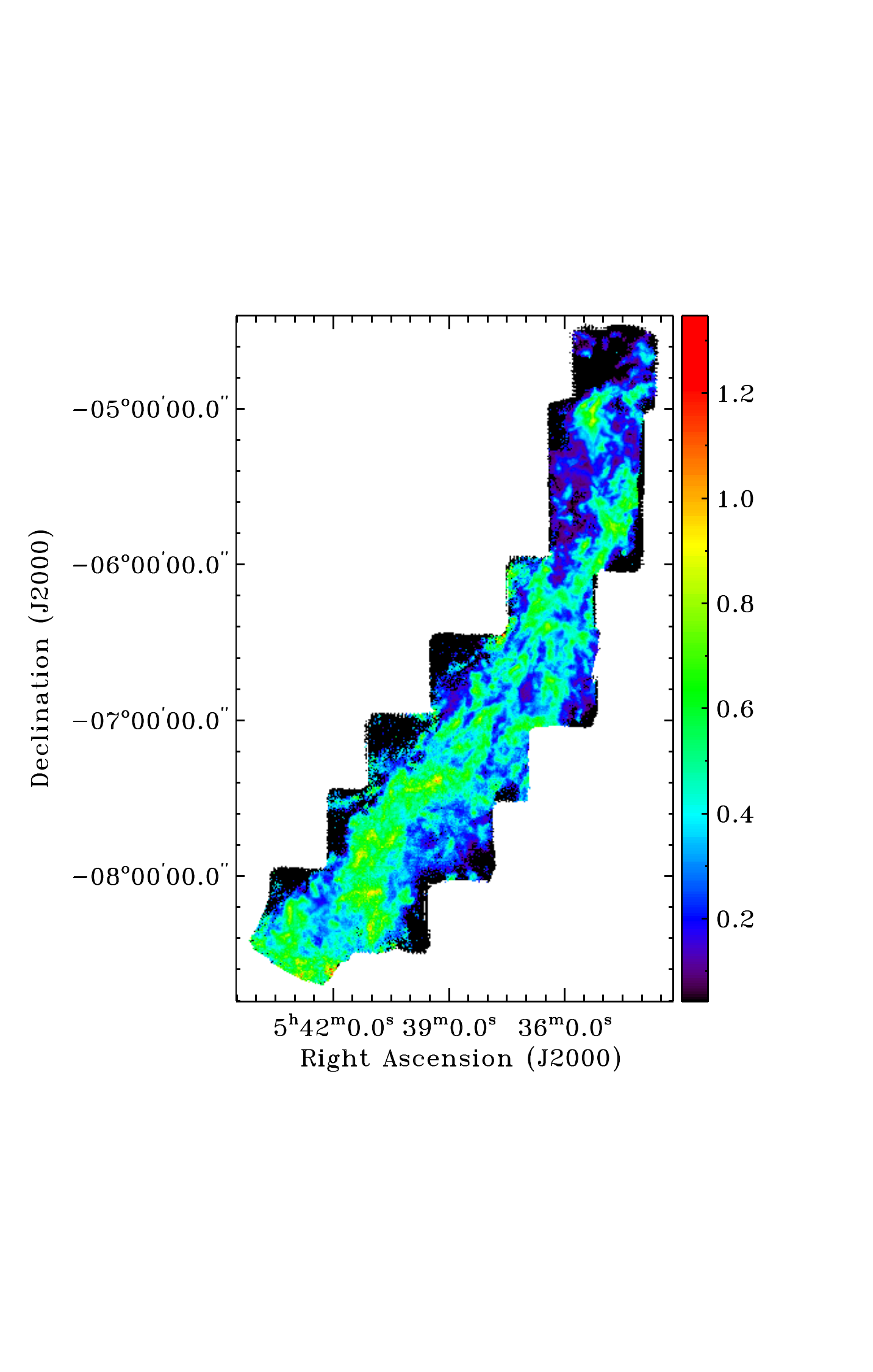}
		\\(b)
	\end{minipage}
	\caption{(a) Excitation temperature derived from the $^{12}$CO emission. (b) Distribution of peak optical depth of the $^{13}$CO emission.}
	\label{Fig2}
\end{figure}

Figure \ref{Fig2}(a) shows the spatial distribution of the excitation temperature derived using Eq. (\ref{eq:1}). The excitation temperatures of Orion A range from 5 to 107 K. In the ISF and L 1641 N regions, most of the excitation temperatures are above 25 K. In the OMC 1 region, which lies behind the Orion Nebula, the temperatures are $\sim$ 50 K on average and reach the maximum 107 K at the position of Orion-KL. Comparatively, the southern part of the GMC is much colder except for the south-eastern tail. The distribution of the $^{13}$CO optical depth is shown in Figure \ref{Fig2}(b). The optical depths, $\tau^{13}$, range from 0.04 and to 1.35 with a median value of 0.37, and only 23 pixels have optical depths greater than 1, indicating the $^{13}$CO emission in the region is mostly optically thin. 


\begin{figure} [!htb]
	\centering
	\begin{minipage}[t]{0.49\linewidth}
		\centering
		\includegraphics[trim = 0cm 5cm 0.5cm 4.5cm, width=\linewidth, clip]{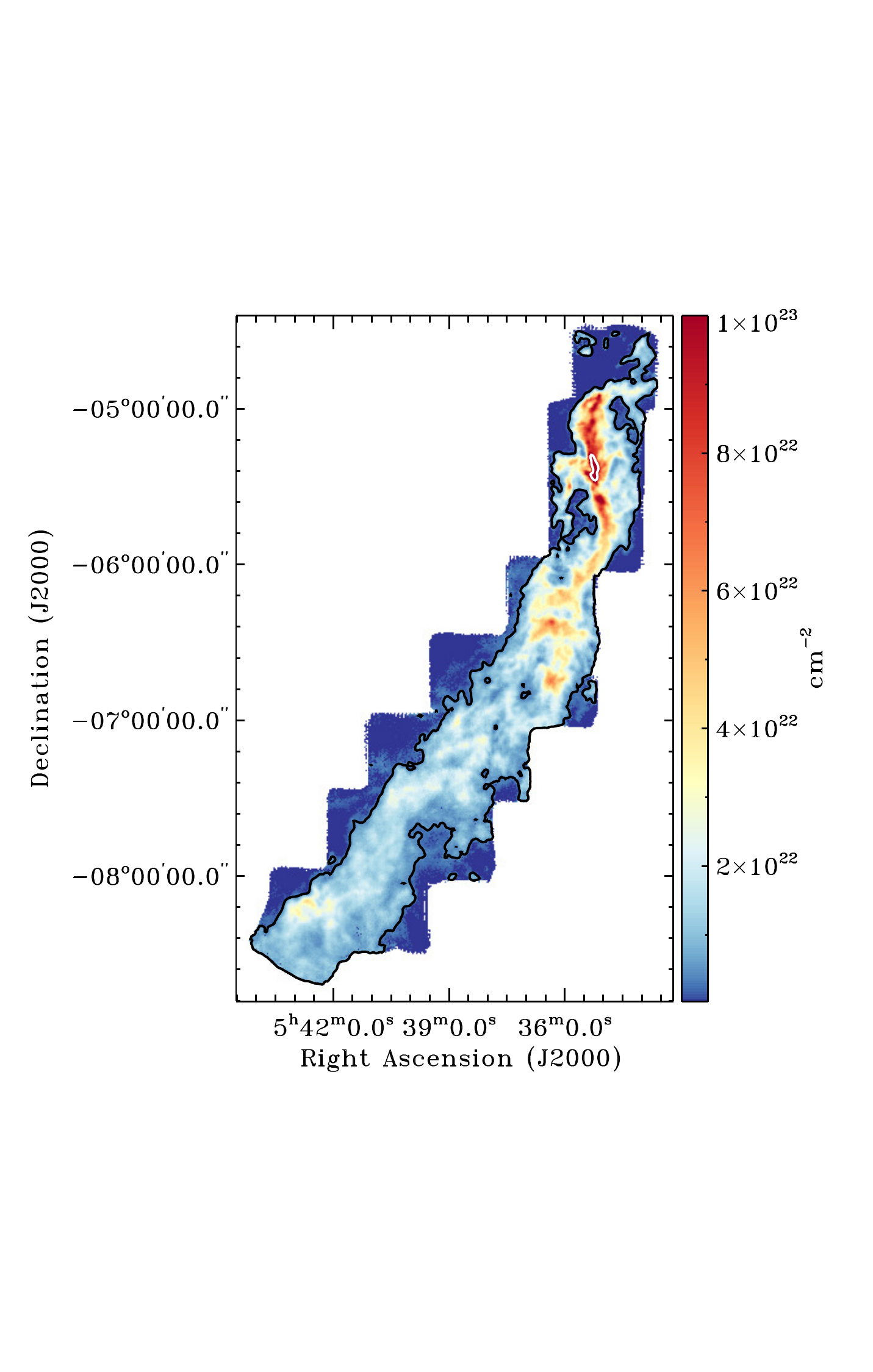}
		\\(a)
	\end{minipage}
	\begin{minipage}[t]{0.5\textwidth}
		\centering
		\includegraphics[trim = 0cm 0cm 0cm 0cm, width=\linewidth, clip]{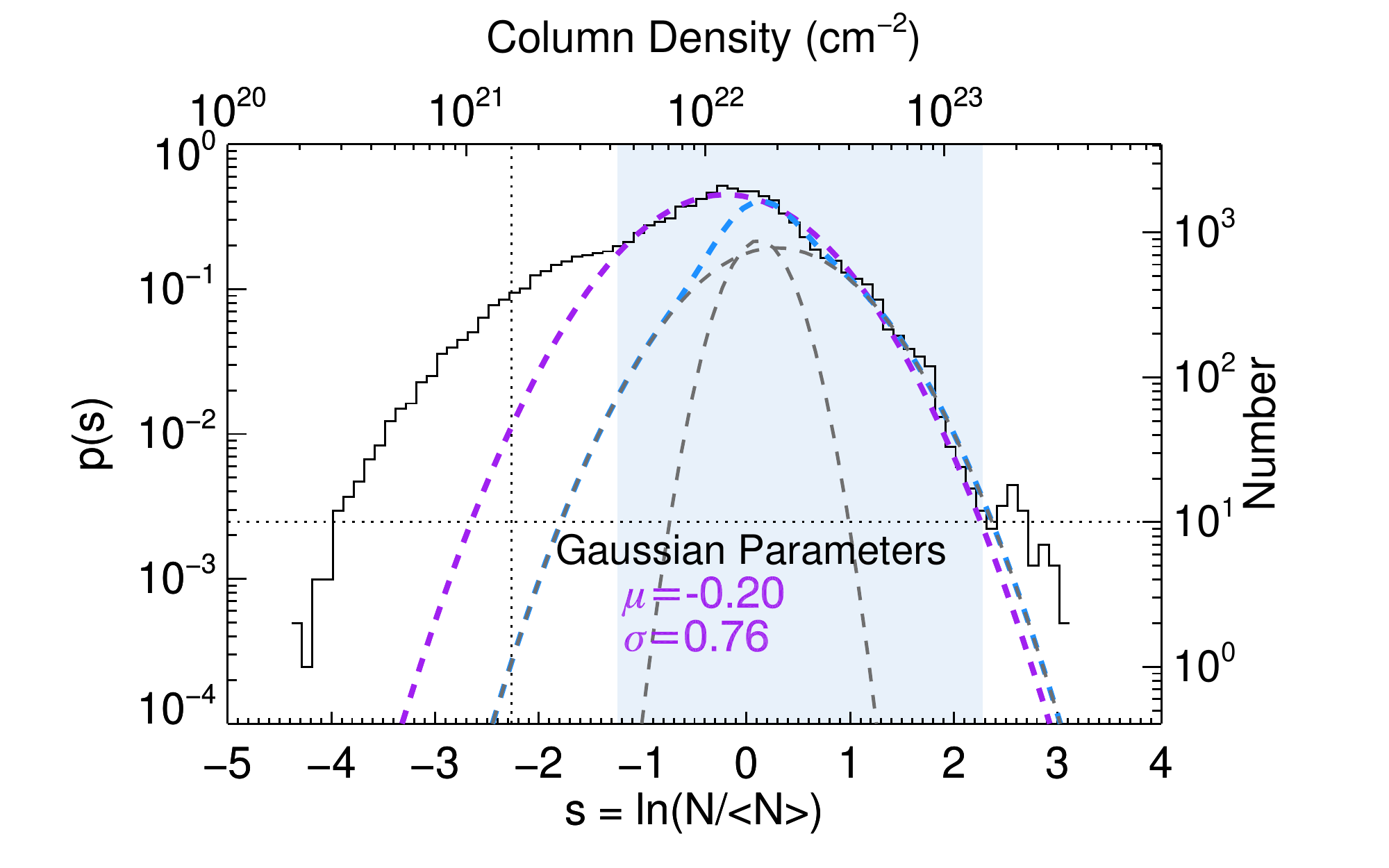}
		\\(b)
	\end{minipage}
	\caption{(a) Spatial distribution of H$_2$ column density of the Orion A GMC. The black contours and the white contours at the middle of ISF correspond to the lower and higher critical value 4$\times$10$^{\rm21}$ cm$^{\rm -2}$ and 1.5$\times$10$^{\rm23}$ cm$^{\rm -2}$, respectively, at which the N-PDF departures from lognormal distribution. (b) N-PDF of H$_2$ column density of the Orion A GMC. The purple dashed line delineates the fitted Gaussian function of $s = ln(N/<N>)$ within the fitting range from $N_{H_2} = 4.4\times10^{21}$ cm$^{-2}$ to $N_{H_2} = 1.4\times10^{23}$ cm$^{-2}$, which is indicated with light blue shadow. The vertical and horizontal dashed lines mark the median detection limit and the level at which the number of pixels in a bin is ten. The fitted parameters $\mu$ and $\sigma$ of the Gaussian function are indicated in purple below the horizontal dashed line. The grey dashed lines are the N-PDFs of the dendrogram structures 5 (narrow) and 11 (wide), respectively, and the blue dashed line is the summation N-PDF of the two structures. For structures 5 and 11, see discussions in Section \ref{sec:4.3}.}
	\label{Fig3}
\end{figure}

\subsection{N-PDF of the Entire Orion A GMC}\label{sec:3.3}
\subsubsection{Overall properties of the H$_2$ column density distribution}\label{sec:3.3.1}
The spatial distribution of the H$_2$ column density of Orion A GMC is displayed in Figure \ref{Fig3}(a). The corresponding pixel-by-pixel statistics is given in \ref{Fig3}(b). The median detection limit based on the selection criteria described in Section \ref{sec:3.1} is 1.5$\times$10$^{21}$ cm$^{-2}$. As shown in Figure \ref{Fig3}(a), the northern ISF and the L 1641 N regions contain more high-density gas than the southern regions. In Figure \ref{Fig3}(b), we totally use 40592 pixels for the statistics and the mean H$_2$ column density is about 1.5$\times$10$^{22}$ cm$^{-2}$, which is much higher than the threshold of star formation of 6.3$\times$10$^{21}$ cm$^{-2}$ \citep{Johnstone2004, Lada2010, Kainulainen2014}. The highest column density is associated with the Orion-KL region, up to a value of $\sim$3$\times$10$^{\rm 23}$cm$^{\rm -2}$. The column densities derived with CO line emission in the Orion A GMC in this work are higher than that derived with the dust tracer \citep{Stutz2015} and are also higher than the column densities derived with CO in other nearby star-forming regions, like the Perseus molecular cloud \citep{Goodman2009a, Kainulainen2009}.  

Mathematically, if a variable N follows a log-normal distribution, its natural logarithm distribution is normal. The log-normal probability density function has the form
\begin{equation}
p(N) = \frac{1}{N}\frac{1}{\sigma\sqrt{2\pi}}e^{-\frac{(lnN - \mu)^2}{2\sigma^2}}
\label{eq:LG}
\end{equation}   
where $\mu$ and $\sigma$ are the mean and the dispersion of $ln N$, respectively. In the histogram in Figure \ref{Fig3}(b), the H$_2$ column density N is normalized by its mean, i.e. the statistical variable is $s = ln(N/<N>)$. As shown in Figure \ref{Fig3}(b), $s$ can be well fitted with a Gaussian function rather than a power-law function in the column density range from 4$\times$10$^{21}$ cm$^{-2}$ to 1.5$\times$10$^{23}$ cm$^{-2}$. In the log-log space, the log-normal function is a parabolic curve, as shown by the purple dashed line in Figure \ref{Fig3}(b). Beyond the fitting range, the N-PDF deviates from the log-normal distribution at both the low-density and the high-density ends. The low-density deviation mostly corresponds to the diffuse gas at the edge of the cloud, which is outside the black contour in Figure \ref{Fig3}(a), while the high-density excess is associated with the very active star-forming regions Orion-KL and Orion-Bar.

\subsubsection{The low-density excess above the log-normal distribution}\label{sec:3.3.2}
\begin{figure}[!htb]
	\centering
	\includegraphics[trim = 0cm 1cm 0cm 1.5cm, width=\linewidth, clip]{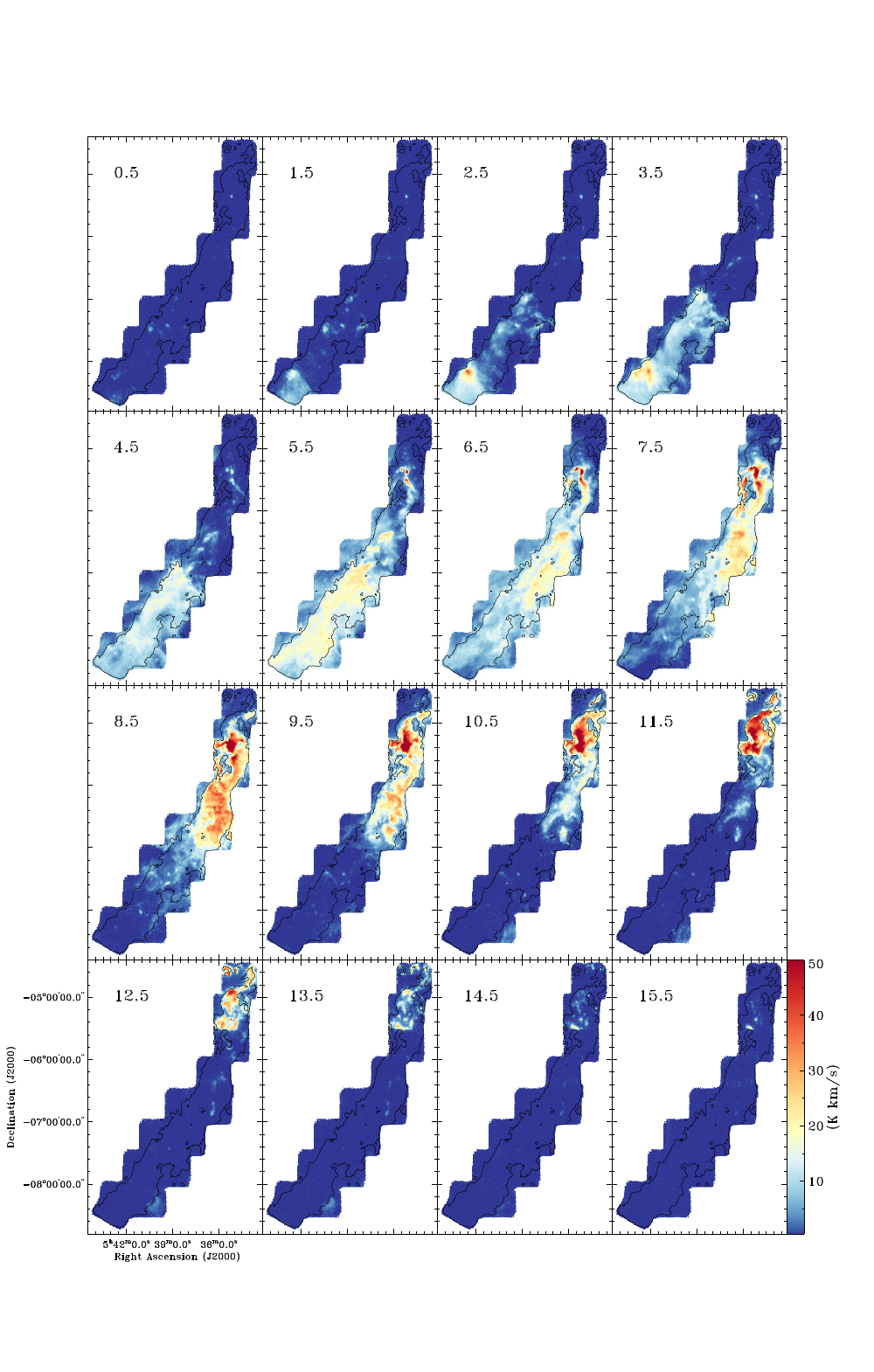}
	\caption{Channel maps of the $^{12}$CO emission with channel width of 1 km s$^{-1}$ in the velocity range from 0 to 16 km s$^{-1}$. The colors start from 1.5 $\sigma_{RMS}$. The black contour in each panel corresponds to N$_{H_2}$ = 4$\times$10$^{21}$cm$^{-2}$. The central velocity of each panel is given in the upper-left corner of each panel.}
	\label{Fig4}
\end{figure} 

\begin{figure}[!htb]
	\centering
	\includegraphics[trim = 1cm 0.5cm 1cm 0cm, width=0.9\linewidth, clip]{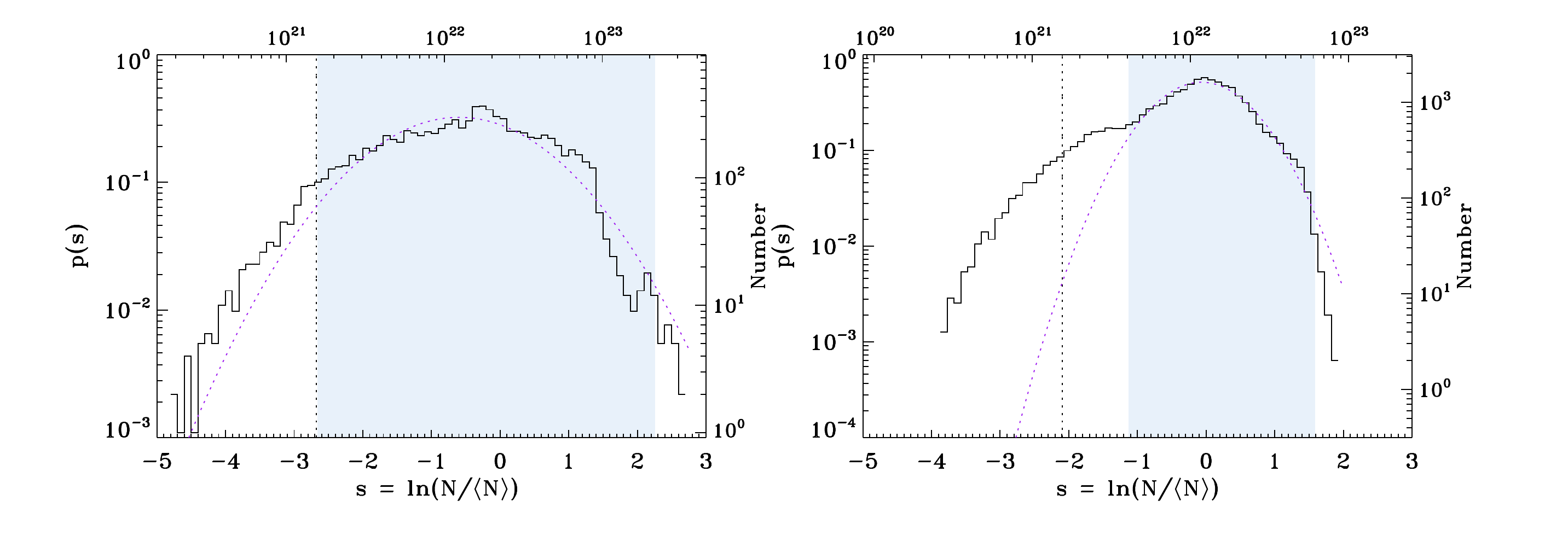}
	\caption{N-PDFs of the two parts of Orion A GMC divided by the declination of $\delta = -5^{\circ}57\arcmin00\arcsec$, overlaid with the fitted log-normal functions as indicated in purple dashed lines. The fitting range in each panel is indicated with light blue shadow.}
	\label{Fig5}
\end{figure}

\begin{figure}[!htb]
	\centering
	\includegraphics[trim = 0cm 0cm 0cm 0cm, width=\linewidth, clip]{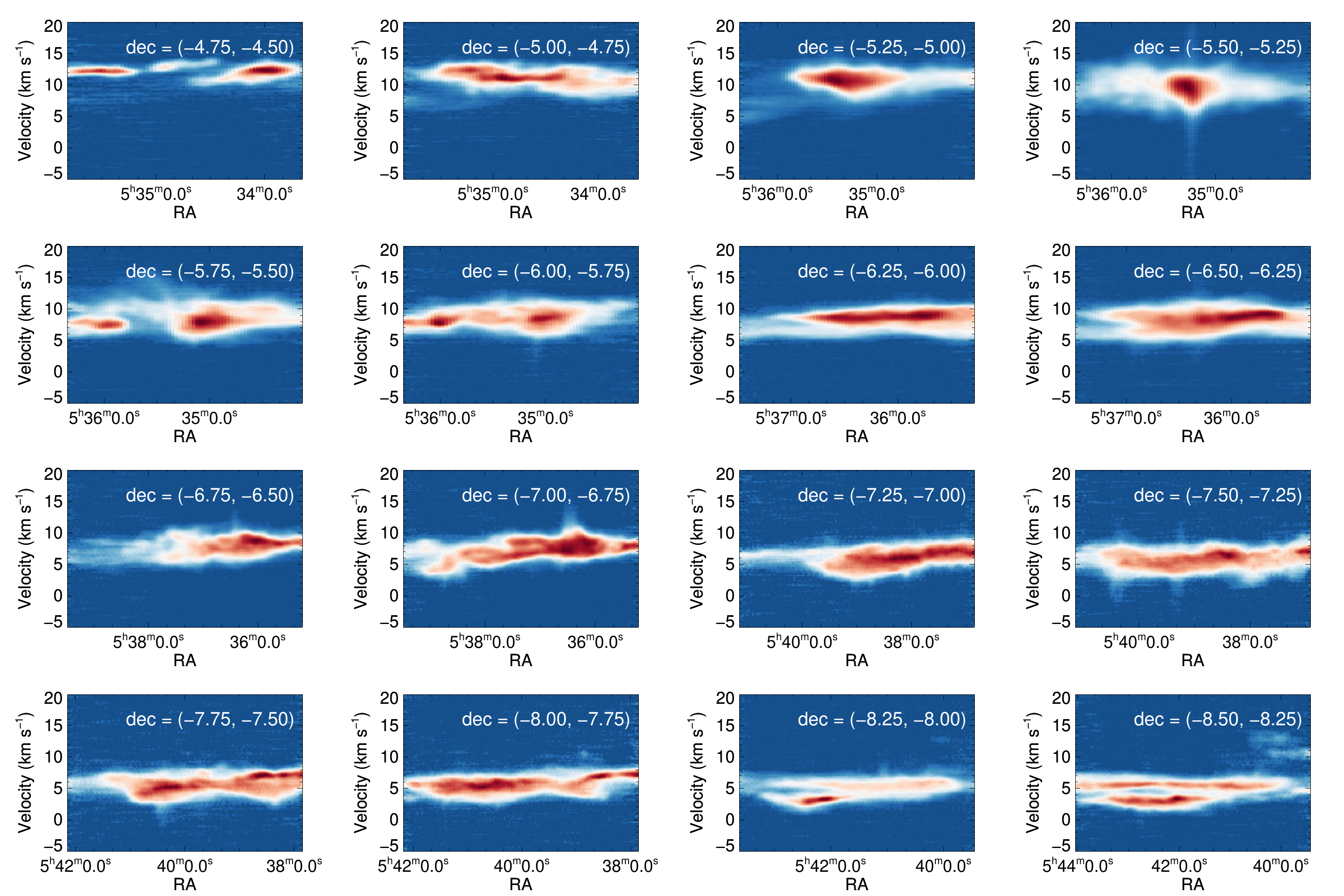}
	\caption{Position-velocity diagrams of the $^{12}$CO emission from declination of -4.5$^{\circ}$ to -8.5$^{\circ}$ in steps of 0.25$^{\circ}$. The colors indicate the integrated intensity in units of K arcdeg from 2 sigma to the maximum of each panel. The declination interval of each panel is shown in white in the upper-right corner.}
	\label{Fig6}
\end{figure}
In Figure \ref{Fig3}(b), the low-density deviation from the log-normal distribution of the N-PDF is quite obvious and is different from the previous results of the Orion A region using either dust or gas tracers \citep{Berne2014, Stutz2015}. We present the channel maps of the $^{12}$CO emission in Figure \ref{Fig4}, and that of $^{13}$CO emission in Figure \ref{FigB1}, to investigate which velocity components contribute to the low-density end of the N-PDF. The 4$\times$10$^{\rm 21}$cm$^{\rm -2}$ contour is overlaid in each panel of Figures \ref{Fig4} and \ref{FigB1}. Outside the 4$\times$10$^{\rm 21}$cm$^{\rm -2}$ contour, extended diffuse gas is present in the velocity channels from 5.5 to 8.5 km s$^{-1}$, more apparent in the $^{12}$CO channel maps than the $^{13}$CO channel maps. From Figure \ref{Fig4}, it can be seen that the diffuse gas mainly exists in the region below $\delta = -5^{\circ}57\arcmin00\arcsec$. Dividing the Orion A GMC by the declination of $\delta = -5^{\circ}57\arcmin00\arcsec$, we separately calculated the N-PDFs of the northern and southern parts of the GMC. The results are presented in Figure \ref{Fig5}. The N-PDFs of the two parts are quite different. The excesses above log-normal distribution at the low-density end of the N-PDF is more obvious in the southern portion than in the northern portion of the GMC.

\begin{figure}[!htb]
	\centering
	\includegraphics[trim = 1cm 1cm 1cm 1cm, width=0.6\linewidth, clip]{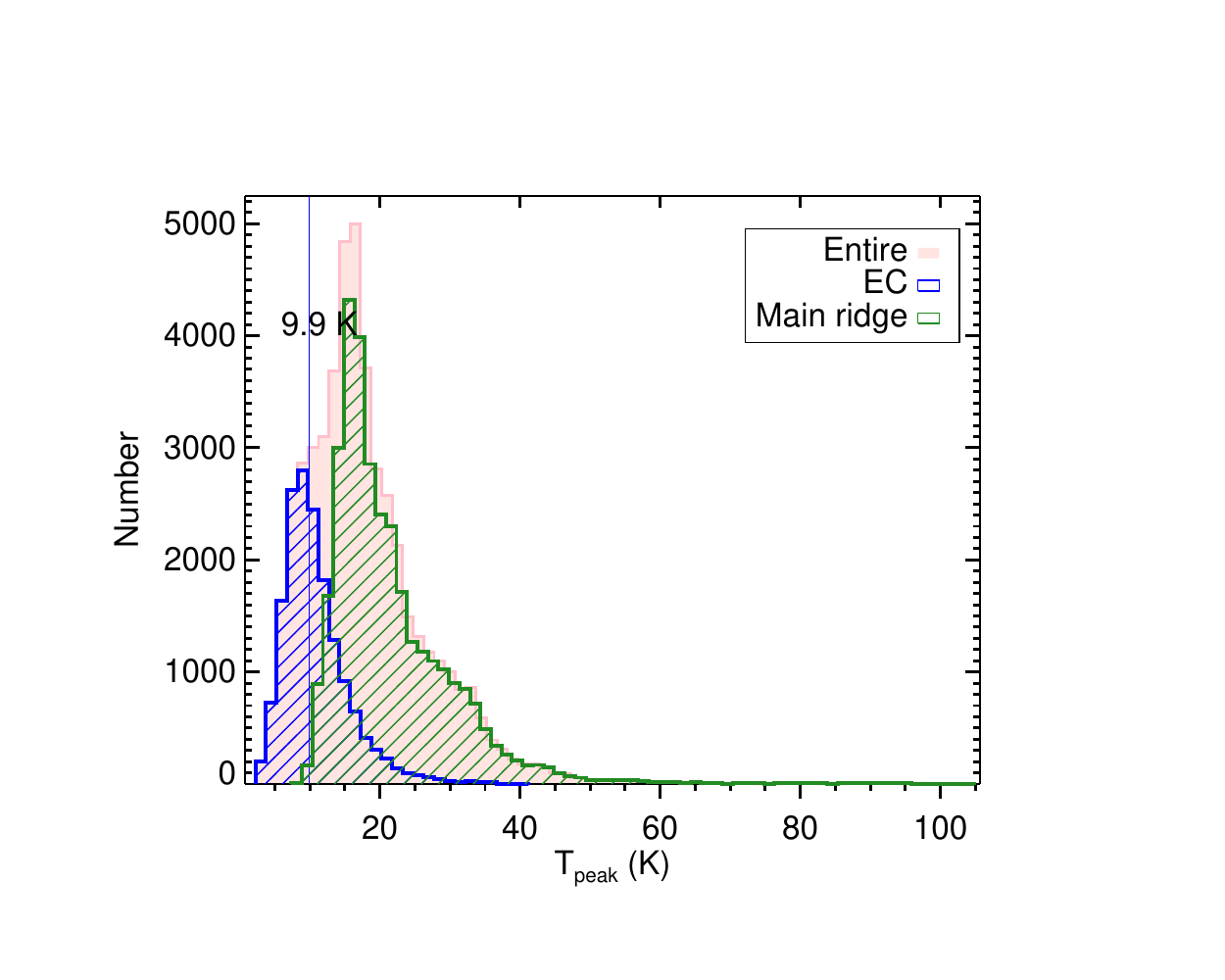}
	\caption{Histograms of the $^{12}$CO peak brightness temperature of the EC counterpart (blue), the main ridge (green), and the entire observed region (pink) of Orion A GMC. The blue vertical line and the black number on that line show the median temperature of the EC counterpart.}
	\label{Fig7}
\end{figure}

Figure \ref{Fig6} presents the position-velocity diagrams of the $^{12}$CO emission at every declination interval of 0.25$^{\circ}$. The P-V diagrams of the $^{13}$CO emission are given in Figure \ref{FigB2} in the Appendix. From Figures. \ref{Fig6} and \ref{FigB2}, it can be seen that there is an apparent velocity gradient from north to south for the main cloud, which is represented by the red color in the panels. However, it can also be seen from the 7th and 8th panels that there is a weak component at 6 km s$^{-1}$, represented by the narrow white strips at the left side of the two panels. This 6 km s$^{-1}$ component is different from the main cloud in terms of the velocity gradient. This component also can be seen in other panels from $-$6\textdegree to $-$7.75\textdegree, although it has similar velocity to the main cloud in regions below $\delta = -$6.5\textdegree, therefore cannot be easily distinguished from the main cloud.    

The 6 km s$^{-1}$ diffuse component is most likely the counterpart of the ``Extended Component (EC)", defined by \cite{Sakamoto1997}. The EC has been fully observed by \cite{Nishimura2015} and they suggest that the EC has a typical velocity of $\sim$ 6 km s$^{-1}$ and is widely distributed to the east of the main ridge of Orion A GMC. We present the histograms of the $^{12}$CO peak brightness temperature for the EC counterpart (N$_{H_2}<4\times10^{21}$ cm$^{-2}$), the main ridge (N$_{H_2}>4\times10^{21}$ cm$^{-2}$), and the entire observed region of Orion A GMC in Figure \ref{Fig7}. The $^{12}$CO emission of the EC counterpart has peak temperatures in the range from 7 to 13 K, which accounts for the low-temperature excess in the histogram of the entire GMC. 

\subsection{N-PDFs in Sub-regions}\label{sec:3.4}
\subsubsection{Calculation of N-PDFs with Star-formation Activities} \label{sec:3.4.1}
\begin{figure}[!htb]
	\centering
	\begin{minipage}[t]{0.55\linewidth}
		\centering
		\includegraphics[trim = 0cm 4.5cm 0cm 5cm, width=\linewidth, clip]{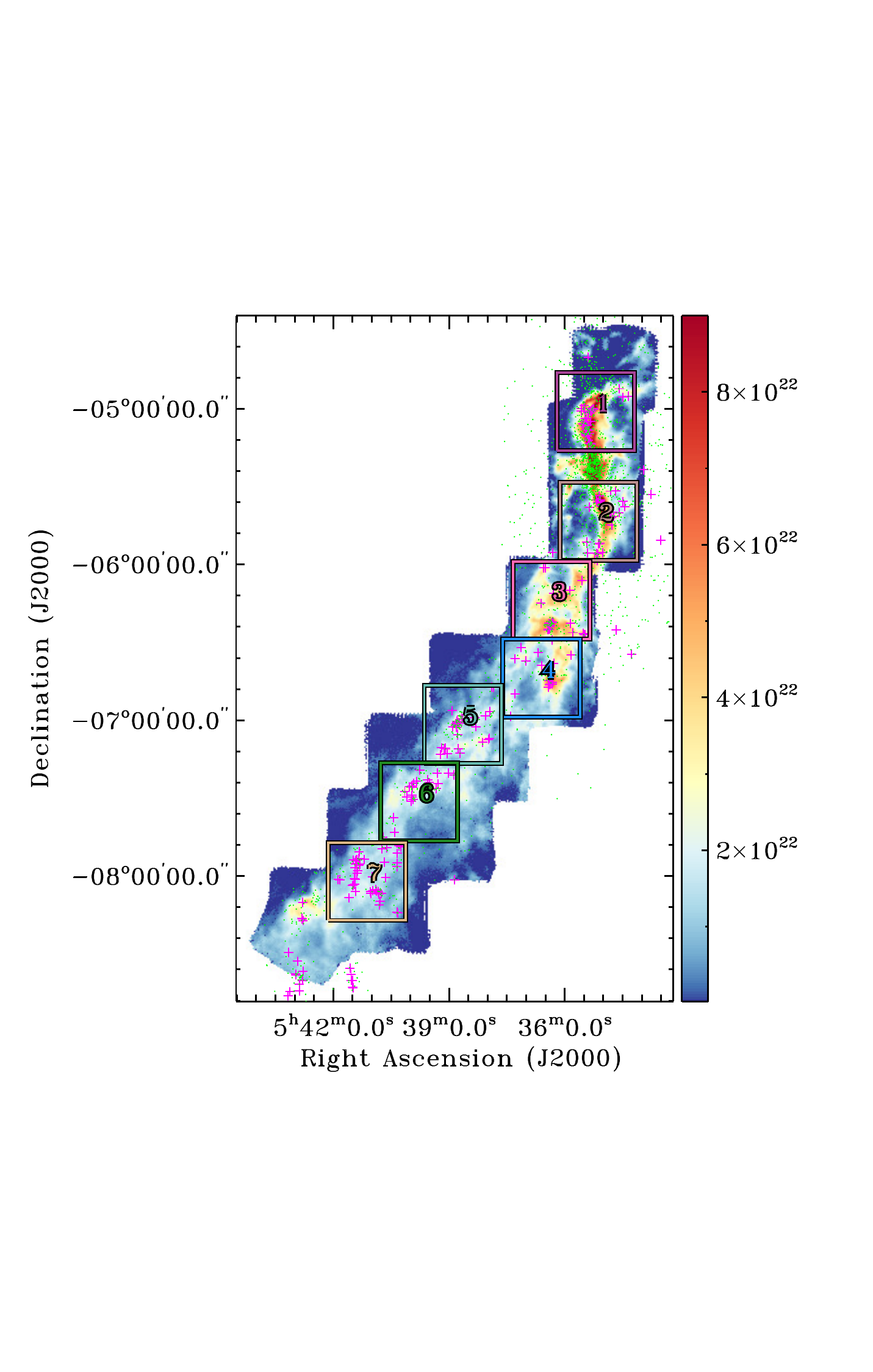} 
		\\(a)
	\end{minipage}
	\begin{minipage}[t]{0.40\textwidth}
		\centering
		\includegraphics[trim = 0.5cm 2.8cm 1cm 2cm, width=0.7\linewidth, clip]{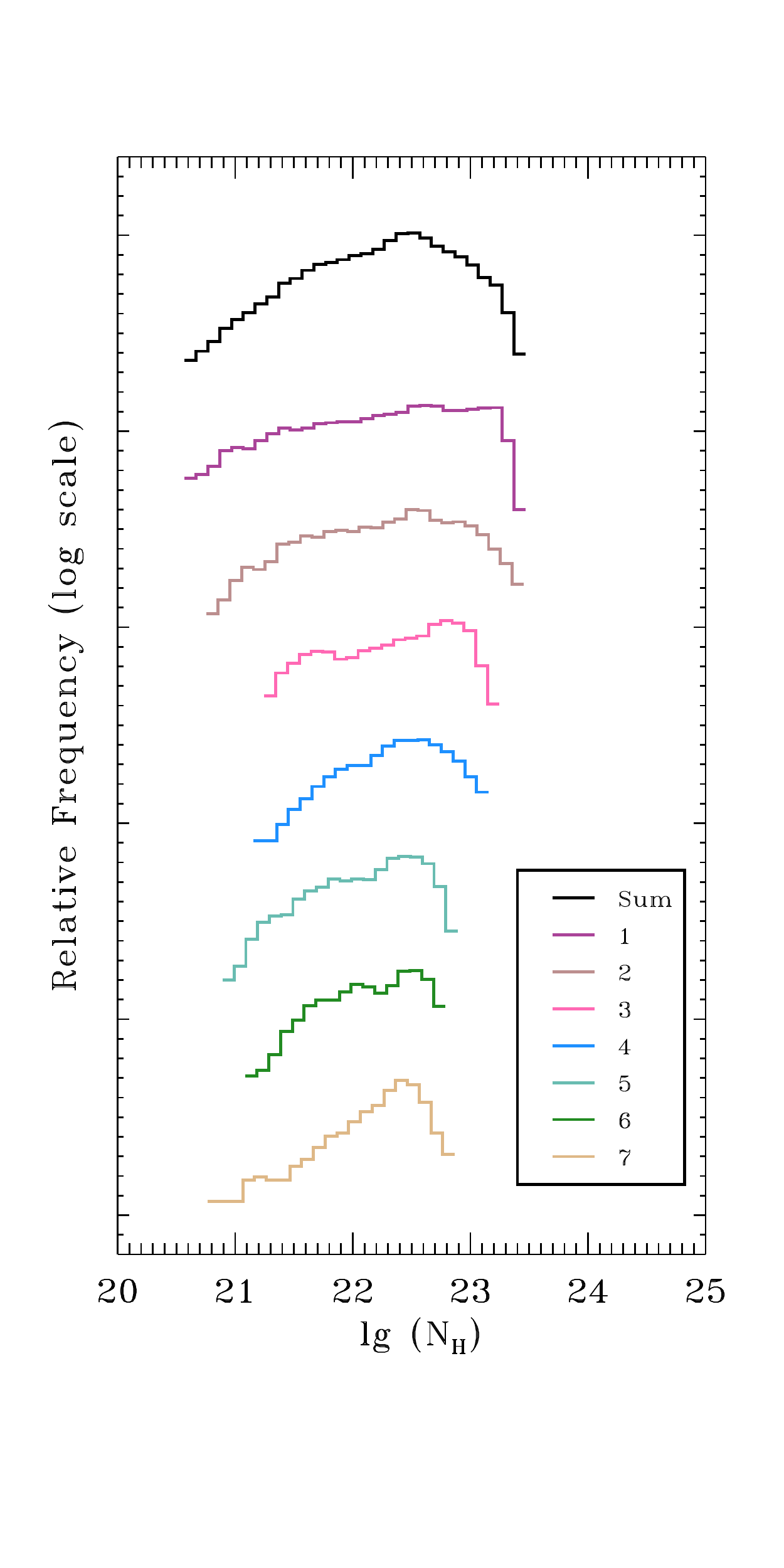}
		\\(b)
	\end{minipage}
	\caption{(a) H$_2$ column density map overlaid with sub-regions 1-7 in S15. The number of each sub-region is indicated in the upper-right corner of the squares. The magenta pluses and the green dots show the positions of the protostars used in S15, and the disk dominated pre-main-sequence stars from the Spitzer catalog \citep{Megeath2012, Megeath2016}, respectively. (b) Hydrogen nucleus N$_H$(2N$_{H_2}$)-PDFs of sub-regions 1-7 and the summation N-PDF of all the sub-regions.}
	\label{Fig8}
\end{figure}

S15 have studied the column density distribution in Orion A using the \emph{Herschel} observations of the dust emission. They used a joint protostar catalog that contains the PACS bright red sources (PBRS) \citep{Stutz2013} and the Class 0, I and Flat-spectrum sources in the Herschel Orion Protostar Survey (HOPS) \citep{Furlan2016} to divide the Orion A GMC into eight 0.5\textdegree$\times$0.5\textdegree sub-regions according to the protostar distribution. They found that the N-PDF of hydrogen nucleus (N$_H$-PDF) derived from dust emission has power-law distribution in each sub-region and the index of the power-law distribution varies with the number fraction of the Class 0 protostars. For direct comparison, we use the first seven sub-regions in S15 to calculate the N$_H$-PDFs except for sub-region 3 that is not fully covered by our observation. The seven sub-regions are indicated with squares in Figure \ref{Fig8}(a) and their corresponding N$_H$-PDFs are presented in Figure \ref{Fig8}(b). The physical parameters of each sub-region, such as the averaged H$_2$ column density, excitation temperature, line width, LTE mass, number of young stellar objects (YSO), and star formation efficiency (SFE), are given in Table \ref{tab1}. In addition to the Class 0 and Class I catalog used by S15, we also included the disk dominated pre-main-sequence stars (Class II sources) from \cite{Megeath2012, Megeath2016} to calculate the SFE in each sub-region. The Class II sources are marked with green dots in Figure \ref{Fig8}(a). 

\begin{table}
	\bc
	\begin{minipage}[]{10cm}
		\caption[]{Physical parameters of each sub-region\label{tab1}}
    \end{minipage}
	\setlength{\tabcolsep}{3.8pt}
	\small
	\begin{tabular}{ccccccccrccc}
		\hline\noalign{\smallskip}
		Region & $\overline N_{H_2}$  & $\mu_s$ & $\sigma_s$ & T$_{ex}$ & T$_{dust}$ &$\sigma_v$ &  M$_{LTE}$ &  n${\rm{_{YSO}}'}$ & n$\rm{_{YSO}}$ & SFE & Mass Fraction \\
		& ($\times$10$^{22}$ cm$^{-2}$) &  & & (K) &(K)& (km s$^{-1}$) & (M$_{\odot}$) & & & &\\
		\hline\noalign{\smallskip}
		  1 &    2.6 &  -0.33 &   0.85 &   29.5 & 27.3 &  0.83 &     5638 &  44 & 420 &     3.59$\%$ &   92.38$\%$\\
		  2 &    2.0 &  -0.28 &   0.76 &   24.7 & 23.0 &1.03 &     5583 &  25 & 408 &     3.52$\%$ &   88.14$\%$\\
		  3 &    2.4 &  -0.18 &   0.65 &   29.2 & 18.2 & 1.07 &     6967 &  22 & 166 &     1.18$\%$ &   92.58$\%$\\
		  4 &    1.8 &  -0.17 &   0.59 &   25.2 & 17.8  & 0.94 &     5265 &  17 &  74 &     0.70$\%$ &   83.34$\%$\\
		  5 &    1.2 &  -0.10 &   0.47 &   20.4 & 16.4 & 0.99 &     3265 &  31 & 127 &     1.91$\%$ &   68.93$\%$\\
		  6 &    1.2 &  -0.12 &   0.51 &   18.1 & 15.4 & 0.96 &     3444 &  26 &  98 &     1.40$\%$ &   72.14$\%$\\
		  7 &    1.3 &  -0.06 &   0.37 &   19.4 & 15.0 & 0.83 &     3634 &  48 & 125 &     1.69$\%$ &   66.97$\%$\\
		\noalign{}\hline
	\end{tabular}
	\ec
	\tablecomments{\textwidth}{Columns 2-4 give the average H$_2$ column density, the mean and dispersion of the natural logarithm of the normalized H$_2$ column density (N$_{H_2} > 4\times10^{21}$ cm$^{-2}$) of each sub-region. Columns 5-8 give the median excitation temperature, median dust temperature from the \emph{Herschel-Planck} temperature map \citep{Lombardi2014}, median velocity dispersion of the $^{13}$CO emission, and the LTE mass of each sub-region. The last four columns give the total number of protostars (Class 0, I and Flat-spectrum sources), the total number of all YSOs (protostars and Class II sources), the SFE, and the dense gas (N$_{H_2}>$1.25$\times$10$^{22}$ cm$^{-2}$) mass fraction of each sub-region. The information of the protostars are from the PBRS and HOPS catalogs \citep{Stutz2013, Furlan2016}, while the information of Class II sources is from the Spitzer catalog \citep{Megeath2012, Megeath2016}. The numbers of the protostars and YSOs are counted within the boundaries of each sub-region, which are the boxes in Figure \ref{Fig8}. The SFE is derived using the assumption that the average YSO mass is 0.5 M$_{\odot}$.} 
\end{table}

\begin{figure}[htb!]
	\centering
	\begin{minipage}[t]{0.45\linewidth}
		\centering
		\includegraphics[trim = 0cm 1.5cm 1cm 1cm, width=\linewidth, clip]{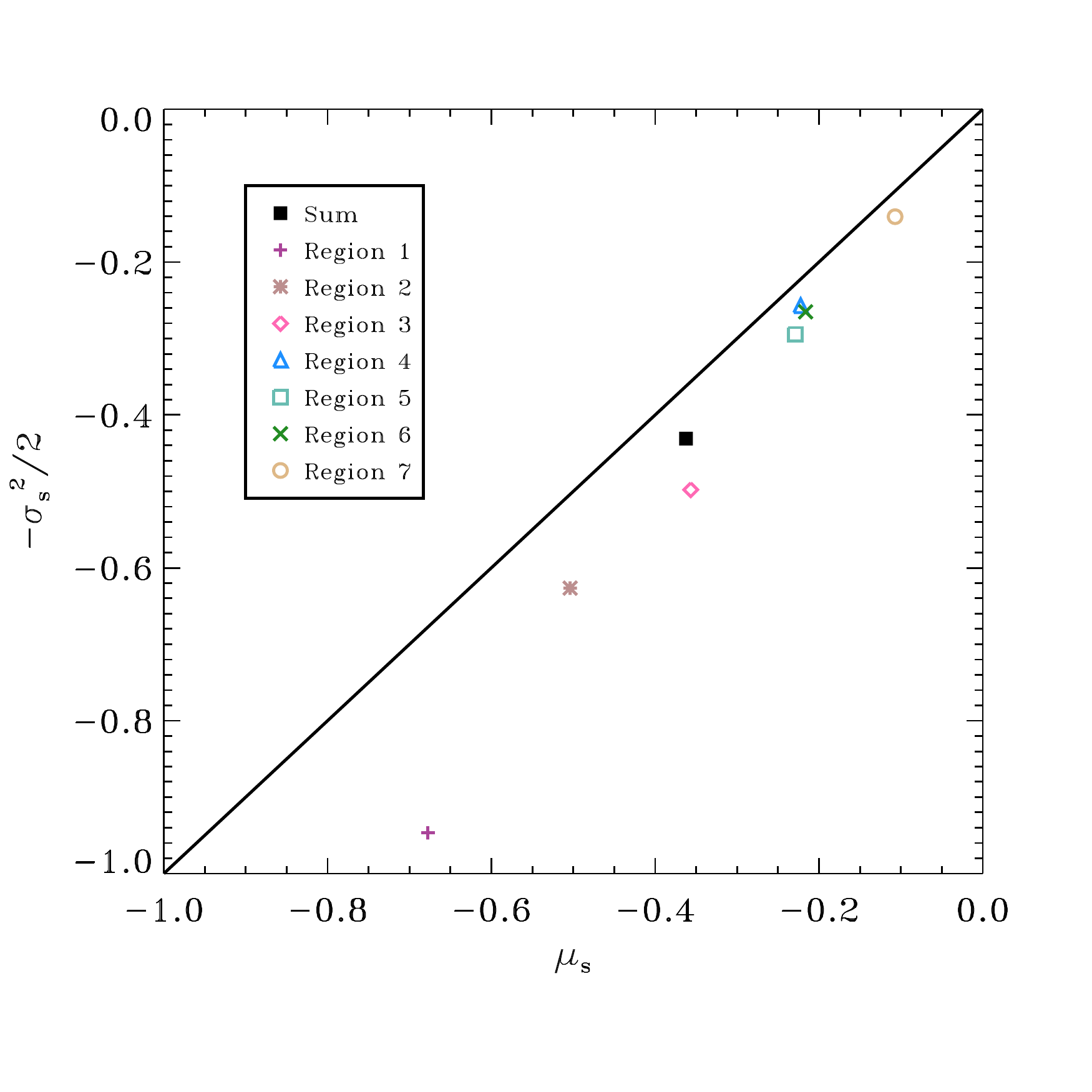}
		\\(a)
	\end{minipage}
	\begin{minipage}[t]{0.45\textwidth}
		\centering
		\includegraphics[trim = 0cm 1.5cm 1cm 1cm, width=\linewidth, clip]{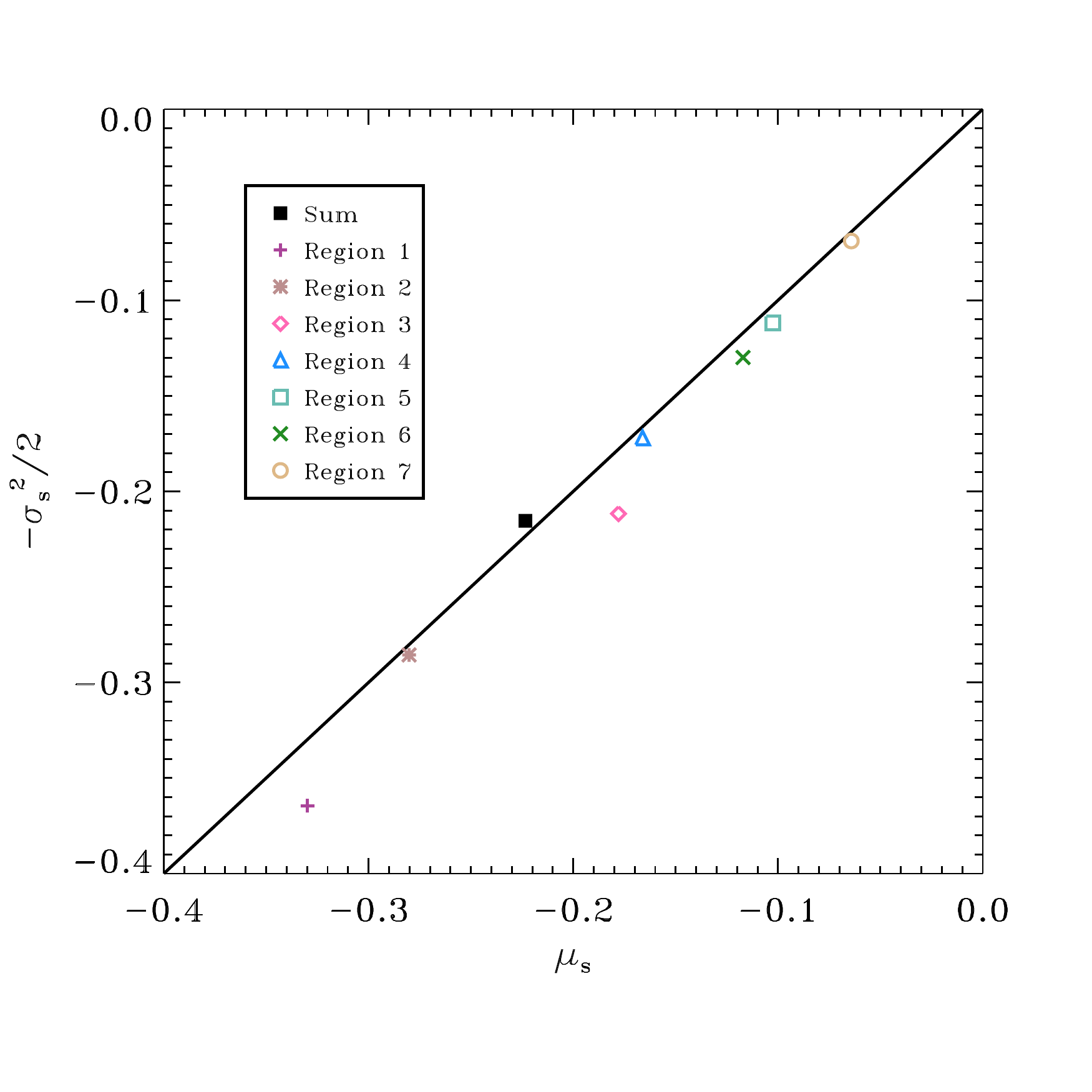}
		\\(b)
	\end{minipage}
	\caption{(a) Relationship between the mean and the dispersion of the logarithmic normalized column densities in the sub-regions and (b) the same relationship but excluding the positions where N$_{H_2}$ is lower than 4$\times$10$^{21}$ cm$^{-2}$. The solid diagonal in each panel represents the relation $\mu_s = -\sigma_s^2/2$.}
	\label{Fig9}
\end{figure}

As shown in Figure \ref{Fig8}(b), although the N$_H$-PDFs of the sub-regions cannot be well fitted with log-normal or power-law functions, it can be seen that the width of the N$_H$-PDF gradually increases from south to north, which resembles the trend of the increasingly flattened power-law N$_H$-PDFs derived using the dust emission (fig. 2 in S15). However, unlike the dust-derived N$_H$-PDFs that can be fitted well by power-law distributions at the high density end, the gas-derived N$_H$-PDFs show quick decline at the high density end, but have relatively extended log-normal-like low-density tails. There is a trend of increasing column density from the south to the north sub-regions. The southernmost sub-regions 5$-$7 have median and maximum H column densities of 2.5$\times$10$^{22}$ and 7.9$\times$10$^{22}$ cm$^{-2}$, respectively, which are a factor of about 1.5 lower than the northern sub-regions. As listed in the last column of Table \ref{tab1}, only $\sim$70$\%$ of the H$_2$ mass is distributed at positions with H$_2$ column density above 1.25$\times$10$^{22}$ cm$^{-2}$ in sub-regions 5$-$7, while this ratio is $\sim$90$\%$ in the northernmost three sub-regions. This difference in column density between the south and north sub-regions may be in part caused by the different extents of CO depletion in these sub-regions. According to chemical evolutionary models of molecular clouds \citep{Bergin1995}, the gas-phase CO abundance is strongly affected by the depletion effect in cold (T$_{dust}\sim$10 K even up to 20 K), dense (n$_{H_2}>$ 10$^4$ cm$^{-3}$), and well-shielded (A$_v\sim$10 mag, corresponds to N$_{H_2} = $ 9.4$\times$10$^{21}$ cm$^{-2}$) \citep{Bohlin1978} regions of molecular clouds, and it is dominated by thermal evaporation in regions with dust temperatures higher than 22 K. It can be seen from Table \ref{tab1} that the dust temperatures in the three southern sub-regions are less than 17 K, indicating that the southern sub-regions are significantly colder than the northern sub-regions. \cite{Ripple2013} have studied the relationship between the $^{13}$CO column density and the extinction in Orion A and found that the $^{13}$CO column density in the southern portion of the GMC, which contains sub-regions 5$-$7 and part of sub-region 4 in this work, no longer increases with extinction when A$_v\sim$ 10 mag. Compared with sub-regions 1$-$4, the lack of high column density gas in sub-regions 5$-$7 is consistent with the scenario that the $^{13}$CO abundance is reduced by depletion in the cold and well-shielded southern sub-regions.     

Whether  an N-PDF is intrinsically a log-normal distribution can be tested with the relationship $\mu_s = -\sigma_s^{2}/2$ resulting from the normalization of the column density \citep{Goodman2009a} (see details in the Appendix). The relationship between the mean and the dispersion of the normalized logarithmic column densities of sub-regions 1$-$7 is shown in Figure \ref{Fig9}(a). The diagonal in Figure \ref{Fig9} represents the relationship $\mu_s = -\sigma_s^{2}/2$ and intrinsic log-normal distributions should fall on this line. As shown in Figure \ref{Fig9}(a), the N-PDFs of the sub-regions more or less deviate from the mean-dispersion relationship for log-normal functions, and the N-PDFs of the southern sub-regions are closer to the line than the northern sub-regions. As discussed in section \ref{sec:3.3.1}, the low-density excess in the N-PDF of the GMC probably originates from the EC counterpart of the Orion A GMC. The $\mu_s-\sigma_s$ relationship when the low-density is excluded shows that the N-PDFs of all sub-regions distribute much closer to the log-normal line (Figure \ref{Fig9}(b)), suggesting intrinsic log-normal distributions of the H$_2$ column density in these sub-regions.  

\subsubsection{The evolutionary trend of the N-PDF} \label{sec:3.4.2}

\begin{figure}[htb!]
	\centering
	\begin{minipage}[t]{0.5\linewidth}
		\centering
		\includegraphics[trim = 0cm 1cm 1cm 1cm, width=\linewidth, clip]{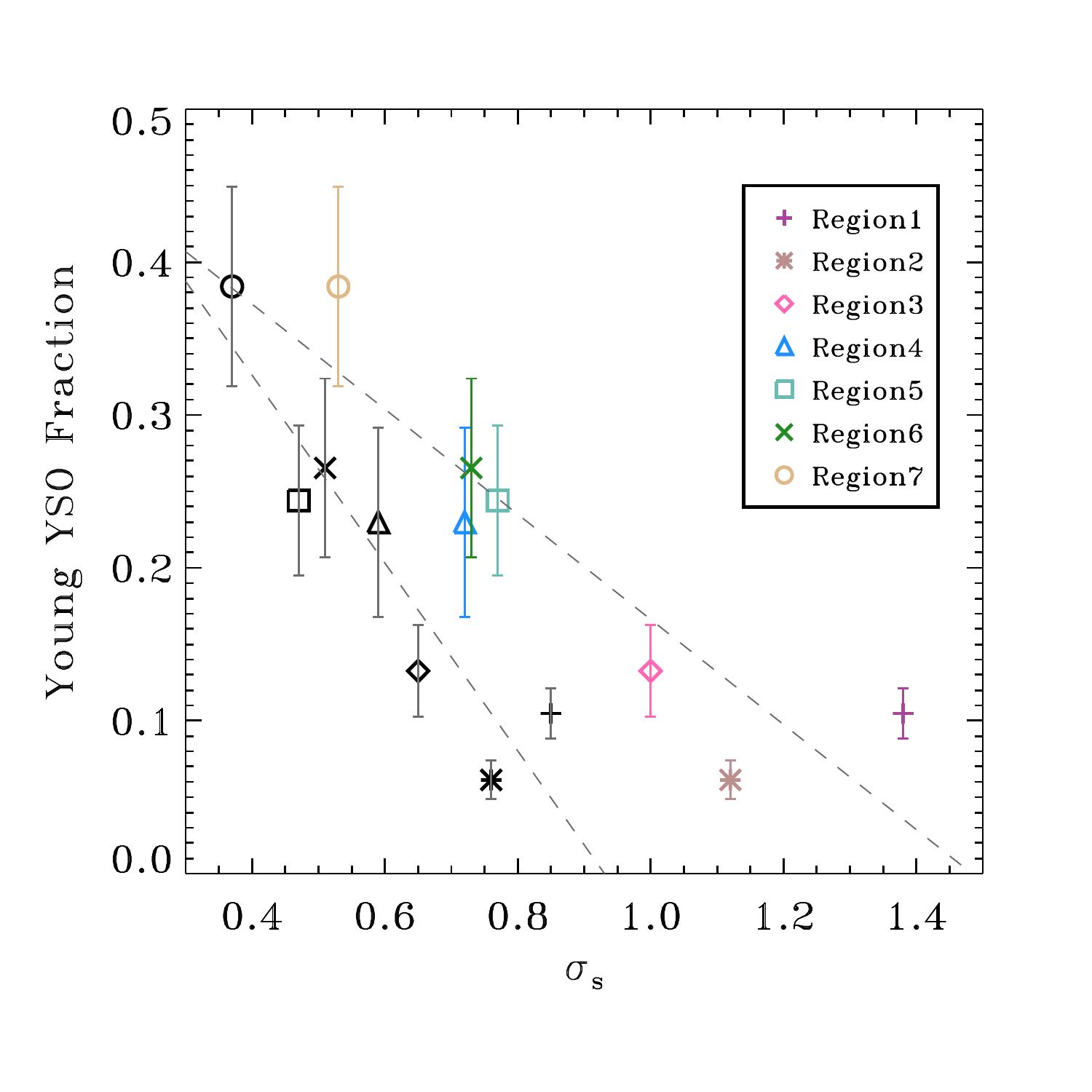}
		\\(a)
	\end{minipage}
	\caption{Relation between the dispersion of the natural logarithmic normalized H$_2$ column density, $\sigma_s$, and the fraction of protostars across the sub-regions. The color and black symbols distinguish between the two cases where the calculation of $\sigma_s$ includes (color) and does not include (black) the positions of N$_{H_2}<$4$\times$10$^{21}$ cm$^{-2}$. The dashed lines give the linear fit to the data points. The Pearson correlation coefficient between $\sigma_s$ and the protostar fraction is $-$0.90 (color) and $-$0.93 (black).}
	\label{Fig10}
\end{figure}

In Table. \ref{tab1}, we summarize the total number of the YSOs and the total number of protostars (Class 0, I, and the Flat-spectrum sources) of each sub-region. The protostars are at an earlier evolutionary stage than the Class II YSOs \citep{Dunham2014}. Figure \ref{Fig10}(a) shows the relationship between the protostar faction and $\sigma_s$ for the sub-regions. If we assume that the Orion A GMC is undergoing star formation with a constant star formation rate, the fraction of protostars is related to the evolutionary status of the molecular cloud. Higher protostar fraction indicates an earlier evolutionary stage. Figure \ref{Fig10}(a) shows that the protostar fraction is anti-correlated with $\sigma_s$, indicating that the column density structure of the molecular clouds in the Orion A region is coupled with the evolutionary stages of star formation, with such a trend that a later evolutionary stage corresponds to a broader column density dispersion.

\section{Hierarchical Column Density Structure} \label{sec:4}  
\subsection{Hierarchical ``Tree" and the Importance of Self-gravity on Different Spatial Scales} \label{sec:4.2}

\begin{figure}[htb!]
	\centering
	\includegraphics[trim = 0cm 0cm 0cm 1cm, width=0.6\linewidth, clip]{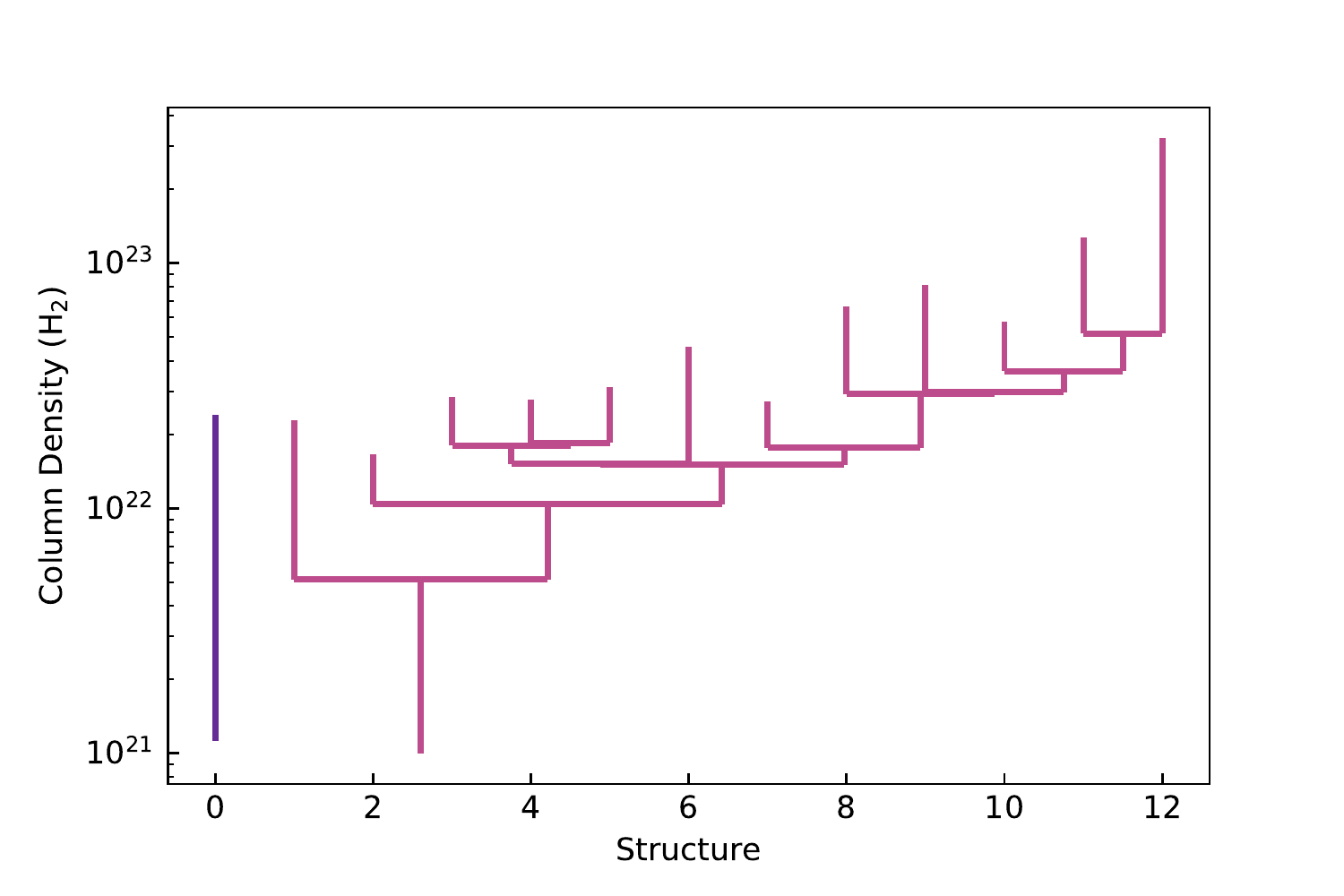}
	\caption{Resulting ``tree" form the DENDROGRAM algorithm, where the values of the x axis have no physical meaning, and different colors correspond to separate ``trunk" structures of the ``tree".}
	\label{Fig11}
\end{figure}

The dendrogram of the tree structure of the Orion A GMC is given in Figure \ref{Fig11}. The observed region contains two ``trunk" structures, as shown in Figure \ref{Fig11}. The right trunk is the main cloud, while the left trunk is the small cloud in the top-left corner of Figure \ref{Fig3}(a). To investigate the properties of structures at various scales, we only use the right trunk in the following analysis. There are 23 individual structures in the right trunk, among which twelve are ``leaves", ten are ``branches", and one is the trunk itself. The sizes of the leaves span a large range and there is no favorable scale for the leaf structure. The leaves contain both large filaments and small clumps. The ``genealogy" diagrams of the spatial distribution of all the structures are given in Figure \ref{Fig12}. As shown in Figure \ref{Fig12}, the ridge of the Orion A GMC is composed of two main structures, branch 11 that contains the entire ISF and the L1641 N regions, and branch 5 that contains the southern tail of Orion A GMC. Comparatively, the density structures are richer in the northern branch than in the southern branch, and there are more levels in branch 11 than branch 5. Downward from branch 3, the northern and southern structures of the GMC merge into one cloud.

The significance of self-gravity for the dynamics of dense cores on scales of $\sim$0.1 pc is well-known \citep{Larson2003, Francesco2007}. However, turbulent simulations without self-gravity successfully yield density and velocity distributions of molecular clouds that resemble observations \citep{Vazquez2001, Padoan2002, Federrath2009}. \cite{Goodman2009b} have made a ``dendrogram" analysis of the L 1448 cloud using $^{13}$CO observations and found that self-gravity is important on the full range of scales of the structures in that cloud, from subparsec to $\sim$ 1 pc. We present the genealogy diagram of the derived virial parameters and N-PDFs of the structures identified in this work in Figure \ref{Fig13}. The panels in Figure \ref{Fig13} that correspond to gravitationally bound structures are filled with light blue. Except for leaf 1 that corresponds to a clump at the southernmost end of the GMC, all structures, on scales from 0.3 pc (leaf 20) to 4.3 pc (trunk 2), are gravitationally bound. This result suggests that self-gravity is also important on much larger scales, i.e., on the clump- to cloud- scales. Our result is consistent with the results of \cite{Goodman2009b} and further highlights the necessity of including self-gravity in simulations of molecular clouds on pc scales.

\subsection{Properties of N-PDFs on Different Spatial Scales}
\label{sec:4.3}
The N-PDFs of the dendrogram structures shown in Figure \ref{Fig13} are uniformly normalized by the mean column density of the Orion A GMC, which is 1.5$\times$10$^{22}$ cm$^{-2}$. We have performed log-normal or power-law fittings on the N-PDFs as appropriate. The results are indicated with dotted and dashed lines in Figure \ref{Fig13}. We note that for most of the dendrogram structures in Figure \ref{Fig13}, the adopted uniform mean column density for the normalization of N-PDF is lower than the actual mean column density of the individual structures, which translates the N-PDFs shown in Figure \ref{Fig13} to the right along the $\ln N/<N>$ axis and therefore increases the value of the fitted $\mu$ parameter. Due to this reason, most of the fitted $\mu$ parameters have positive values.   

\begin{figure}
	\centering
	\includegraphics[trim = 2.3cm 2.3cm 2.3cm 2.3cm, width=\linewidth, clip]{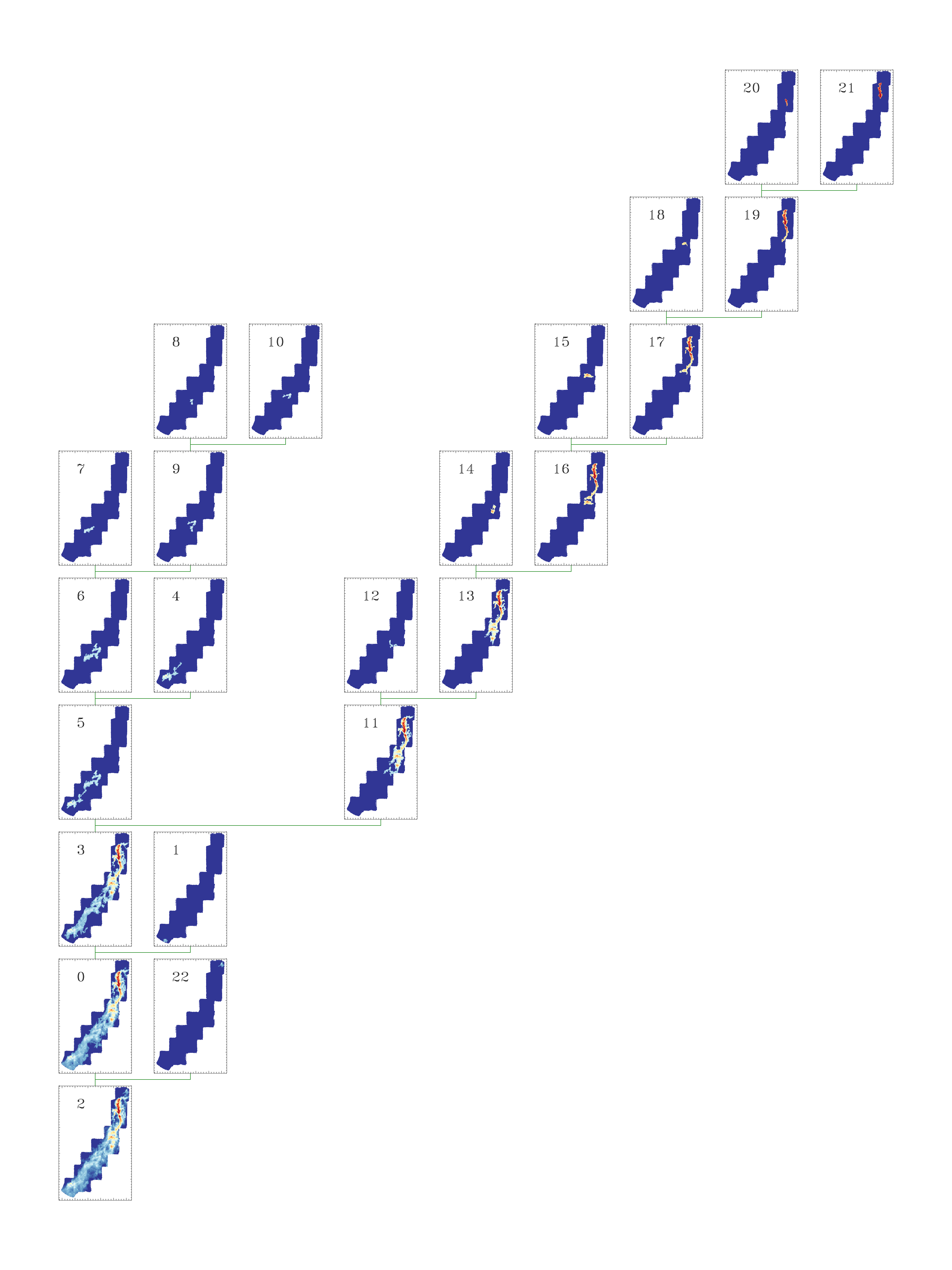}
	\caption{``Genealogy" diagram of the sub-structures in the dendrogram in Figure \ref{Fig11} and their spatial distribution. The number in each panel is the index of the structure in the dendrogram. The green line shows the relationship between the structures.}
	\label{Fig12}
\end{figure}

\begin{figure}
	\centering
	\includegraphics[trim = 1cm 0cm 1cm 1cm, width=1.5\linewidth, clip, angle = -90]{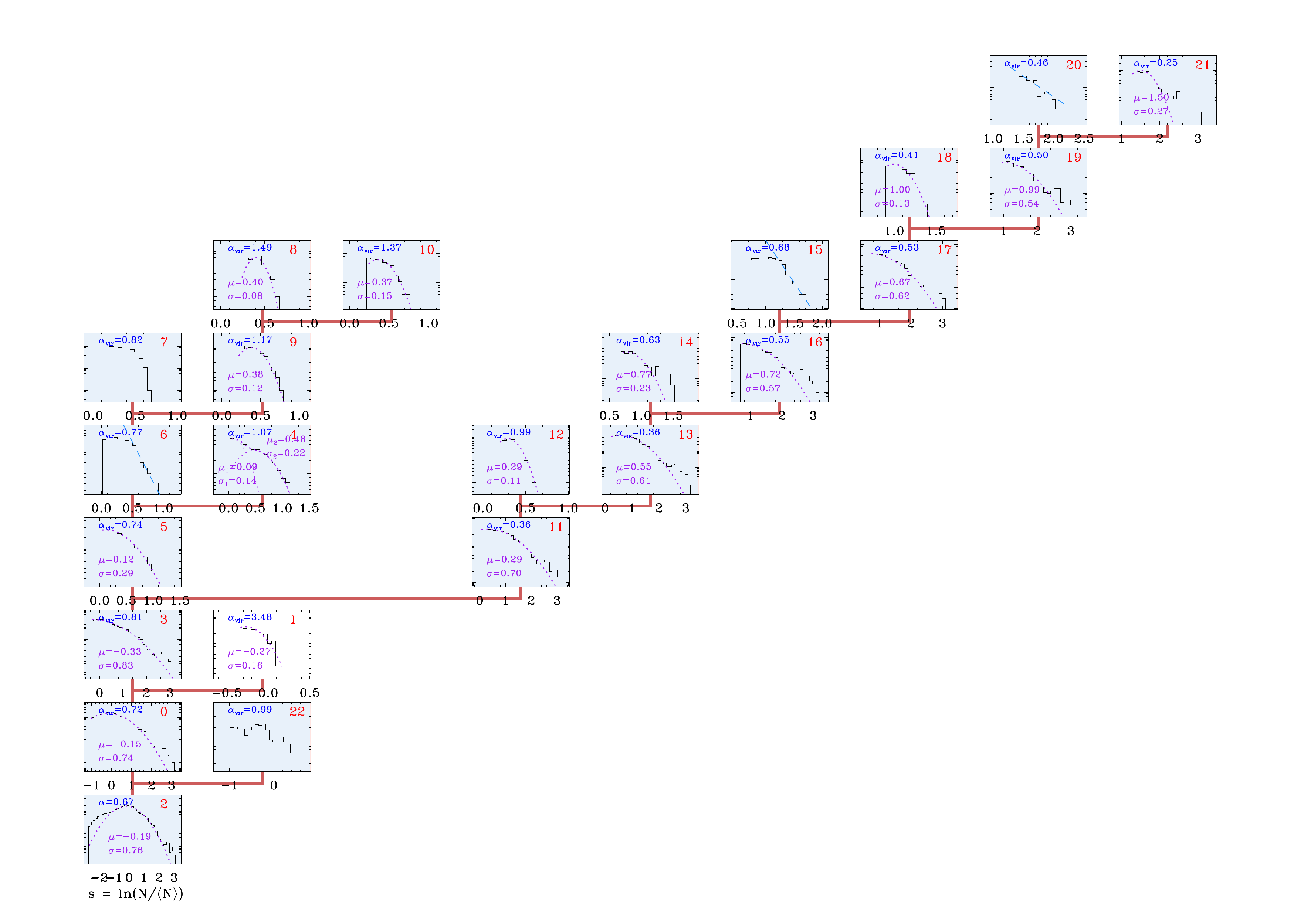}
	\caption{Same configuration as Figure \ref{Fig12}, but with N-PDF rather than the spatial distribution of each structure being presented in each panel. The log-normal and power-law fit to the N-PDFs are indicated with the purple dotted line and the blue dashed line, respectively. The fitted log-normal parameters are given in each panel. The virial parameter of each structure is shown in the upper-left corner in each panel. The panels corresponding to structures with $\alpha < 2$ are filled with light blue.}
	\label{Fig13}
\end{figure}

The leaves of branch 11 can be divided into three categories, leaves 18 and 12 with pure log-normal N-PDFs, leaves 21 and 14 that have log-normal N-PDFs with high-density excesses, and leaves 20 and 15 that have power-law N-PDFs at the high-density end. As shown in Figure \ref{Fig12}, leaf 18 is located to the south of the ISF structure, and corresponds to the center of sub-region 3 in Section \ref{sec:3.4}. A relatively low number of protostars are associated with leaf 18, indicating inactive star formation within leaf 18. Leaf 12 is a relatively faint filament located at the southernmost end of branch 11, and this filament is also quiescent in star formation as indicated by the few associated Class II YSOs in Figure \ref{Fig8}. Leaves 21 and 14 correspond to the OMC-1/2/3 regions, which contain a cluster of intermediate-mass protostars and the NGC 1999 region. The high-density excesses in leaves 21 and 14 are concentrated in regions with s$\sim$ 2.3$-$3.2 ($N_{H_2}\sim1.5-3.7\times10^{23}$ cm$^{-2}$) and s$\sim$ 1.2$-$1.5 ( $N_{H_2}\sim5.0-6.7\times10^{22}$ cm$^{-2}$), respectively. The high-density regions account for 12.8\% of the mass and 4.2\% of the area for leaf 21. For leaf 14, the high-density mass and area fractions are 23.7\% and 16.6\%, respectively. Leaves 20 and 15 correspond to OMC-4, which includes a dense star-forming region and the H II region L 1641 N. Leaf 20 is possibly on a later evolutionary stage than leaf 21, as indicated by the pure power-law PDF and the lower protostar fraction as discussed in Section \ref{sec:3.4.2}. The N-PDFs of branches 19, 17, 16, 13, and 11 can be well-fitted with log-normal functions below $s = 2.3$. The high-density excesses in the N-PDFs of these branches are caused by the pixels of density above $N_{H_2}\sim1.5\times10^{23}$ cm$^{-2}$ which is confined to the Orion KL region in leaf 21 and is shown with the white contour in Figure \ref{Fig3}a. The high-density pixels with $s>2.3$ only account for a small mass and area proportion of the branch N-PDFs below leaf 21. The mass proportion decreases from 12.8\% of leaf 21 to 2.6\% of trunk 2, while the corresponding area proportion decreases from 4.2\% to 0.2\%. Therefore, the majority of the structures of branch 11 are dominated by log-normal N-PDFs and departures from log-normal N-PDFs at high-densities only occur in active star-forming regions.

Except for leaf 7 that corresponds to the H II region L 1641 S, all leaves in branch 5 have log-normal N-PDFs without high-density excesses. The N-PDF of leaf 4 is well fitted with double log-normal functions, indicating the possible existence of mixed components along the line-of-sight. We can see from the last two panels in Figure \ref{Fig6} that there are two separate velocity components at the location of leaf 4, one of which is in the range from 5 to 8 km s$^{-1}$, the same as the EC counterpart. However, it is most likely that the velocity separation in the leaf 4 region is a local dynamic property and not caused by the EC gas. The N-PDFs of branches 9 and 5 are fitted with log-normal functions, while the N-PDF of branch 6 that contains the H II region L 1641 S is fitted with power-law at its high density tail. 

Generally, the $N_{H_2}$ dynamical ranges of the structures above branch 5 are smaller than those of the structures above branch 11. From leaf 21 to branch 11, the $\mu$ of the fitted log-normal function decreases from $\sim$1.5 to $\sim$0.3, and the $\sigma_s$ increases from $\sim$0.3 to $\sim$0.70. From leaf 8 to branch 5, the $\mu$ and the $\sigma_s$ of the fitted log-normal functions vary gradually from $\sim$0.4 to $\sim$0.1 and $\sim$0.1 to $\sim$0.3, respectively. The ranges of $\mu$ and $\sigma_s$ of structures in branch 5 are much smaller than those in branch 11, indicating the southern filament (branch 5) of the Orion A GMC is at an earlier evolutionary stage than the northern filament (branch 11), which is consistent with the results in Section \ref{sec:3.4}. We also plot the fitted log-normal N-PDFs of branches 5 and 11, and their summation in Figure \ref{Fig3}. The N-PDFs of the two filaments account for the high-density part of the column density distribution of Orion A, supporting that the Orion A molecular cloud is mainly composed of these two filaments. 

Below branches 11 and 5, there are two leaves and three branches. Leaves 1 and 22 correspond to a southernmost and a northernmost clump, respectively, both inactive in star formation. The N-PDF of leaf 1 is fitted with a log-normal function while the N-PDF of leaf 22 has an irregular shape. Branch 3 is mainly the combination of filaments 5 and 11, and is approximately the main ridge of the Orion A GMC. The N-PDFs of branches 3, 0, and 2 are fitted with log-normal functions and getting closer to the overall N-PDF of the whole cloud successively. The high-density excesses in the N-PDFs of these three branches are caused by leaf 21, while the low-density excess of branch 2 is caused by the EC gas, as discussed in Section \ref{sec:3.3.2}.

In summary, regardless the sub-structure is in the northern or southern part of the GMC, power-laws and deviations from log-normal distributions at the high-density tail of N-PDFs only exist in active star-forming regions. The majority of the structures in Orion A GMC have log-normal dominated N-PDFs on scales from $R\sim0.4$ to 4 pc. The predominant existence of log-normal N-PDFs in structures on such a broad range of scales suggests a dominant role of turbulence on these spatial scales. 

\section{Summary and Conclusion}\label{sec:5}
We have conducted a large-field survey of $^{12}$CO, $^{13}$CO, and C$^{18}$O $J=1-0$ emission toward the Orion A GMC with a sky coverage of $\sim$ 4.4 deg$^{2}$. Using the $^{12}$CO and $^{13}$CO data, we have investigated the N-PDFs of H$_2$ column density of the GMC and compared the N$_{H}$-PDFs with previous results from \emph{Herschel} observations in seven sub-regions. The importance of self-gravity and properties of N-PDFs on different spatial scales are studied using the DENDROGRAM method. The main results are summarized as follows. 
\begin{enumerate}
\item The H$_2$ column density N-PDF for the entire GMC is fitted with a log-normal function in the range from $\sim$4$\times10^{21}$ to $\sim$1.5$\times10^{23}$ cm$^{-2}$, with excesses at both the low-density and the high-density ends. The excess at the high-density end corresponds to the Orion KL region, and the excess at the low-density end is possibly caused by an extended and low-temperature component ($\sim$10 K) with velocities in the range of 5$-$8 km s$^{-1}$. 
\item  To compare with the results from \emph{Herschel} observations, we divided the Orion A GMC into seven sub-regions. The N-PDFs of the sub-regions are irregular in shape, but are log-normal distributed at the high column density end. Compared with the northern sub-regions, the three southern sub-regions have less high-density gas, which may be in part caused by the CO depletion effect. The H$_2$ column density structure is coupled with the evolutionary stage of clouds in Orion A GMC. Broader column density dispersions correspond to later evolutionary stages.
\item In terms of structure hierarchy, Orion A GMC is mainly composed of two filamentary structures located in the north and south, respectively. All the structures in the dendrogram tree except leaf 1 are gravitationally bound across spatial scales from 0.3 to 4.3 pc. Although power-laws and departures from log-normal distributions exist in N-PDFs of structures in five active star-forming regions mostly with scales $<0.6$ pc, the N-PDFs of structures in the Orion A GMC are predominantly log-normal on scales from $R\sim0.4$ to 4 pc. 
\end{enumerate}

\begin{acknowledgements}		
We thank the PMO-13.7 m telescope staffs for their supports during the observation. MWISP project is supported by National Key R$\&$D Program of China under grant 2017YFA0402701 and Key Research Program of Frontier Sciences of CAS under grant QYZDJ-SSW-SLH047. Y. Ma acknowledges supports by NSFC grants 11503086 and 11503087. This work makes use of the SIMBAD database, operated at CDS, Strasbourg, France.	
\end{acknowledgements}
\clearpage
\appendix    
\section{The relationship between the mean and the dispersion of the logarithmic normalized column density} \label{sec:A}
As the definition of the log-normal distribution introduced in Eq. (\ref{eq:LG}) in Section \ref{sec:3.3}, the mathematical expectation of N is 
\begin{equation}
\centering
E(N) = \int_{0}^{+\infty}Nf(N)dN = \int_{0}^{+\infty}\frac{1}{\sigma\sqrt{2\pi}}e^{-\frac{(lnN-\mu)^2}{2\sigma^2}}dN.
\label{eq:A1}
\end{equation}
where $N$, $\sigma$, and $\mu$ has the same meaning as in Eq.(\ref{eq:LG}). If we let $t = (ln\ N-\mu)/(\sqrt{2}\sigma)$, then $N = exp(\sqrt{2}\sigma t+\mu)$  and $dN = \sqrt{2}\sigma exp(\sqrt{2}\sigma t+\mu)\ dt$. The expectation can be written as 
\begin{equation}
E(N) = \int_{-\infty}^{+\infty}\frac{1}{\sqrt{\pi}}e^{-t^2+\sqrt{2}\sigma t+\mu}dt = \frac{1}{\sqrt{\pi}}\int_{-\infty}^{+\infty}e^{-(t-\frac{\sqrt{2}}{2}\sigma)^2+\mu+\frac{\sigma^2}{2}}dt.
\label{eq:A2}
\end{equation}
Again, after make the substitution of $u = t-\sqrt{2}/2\sigma$ and make use of the integral formula $\int_{-\infty}^{+\infty}e^{ax^2}dx = \sqrt{\pi/a}$ the above equation is finally converted to 
\begin{equation}
E(N) = \frac{1}{\sqrt{\pi}}e^{\mu+\frac{\sigma^2}{2}}\int_{-\infty}^{+\infty}e^{-u^2}du = e^{\mu+\frac{\sigma^2}{2}}.
\label{eq:A3}
\end{equation}
Because of the normalization when calculating the N-PDF in this work, the mathematical expectation of N should be unity which means $\mu+\sigma^2/2 = 0$, i.e., $\mu = -\sigma^2/2$.
             
\section{Figures}
\begin{figure}[bht!]
	\centering
	\includegraphics[trim = 0cm 1cm 0cm 1.5cm, width=\linewidth, clip]{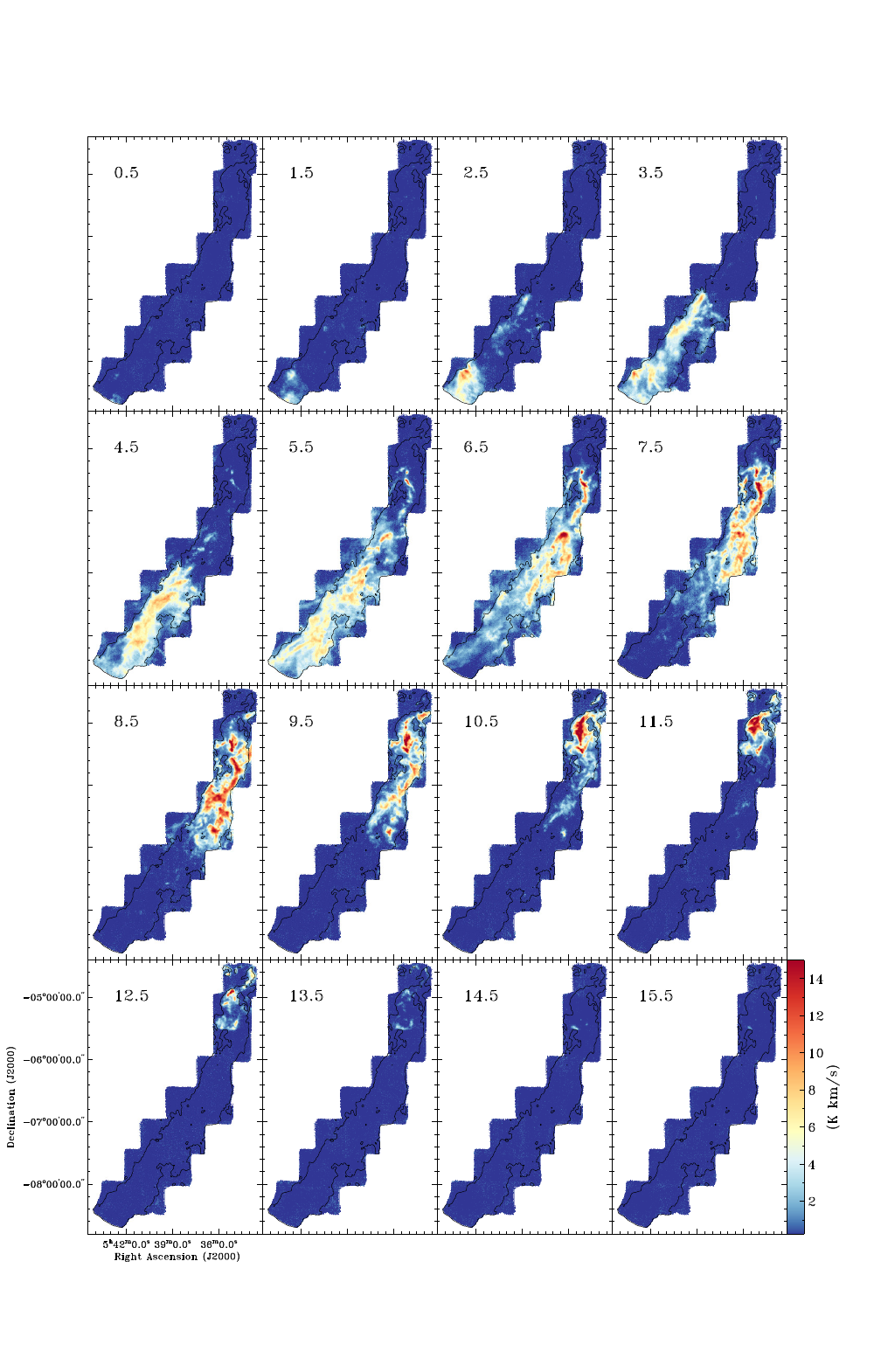}
	\caption{Same as Figure \ref{Fig4} but for the $^{13}$CO emission.}
	\label{FigB1}
\end{figure}

\begin{figure}
	\centering
	\includegraphics[trim = 0cm 0cm 0cm 0cm, width=\linewidth, clip]{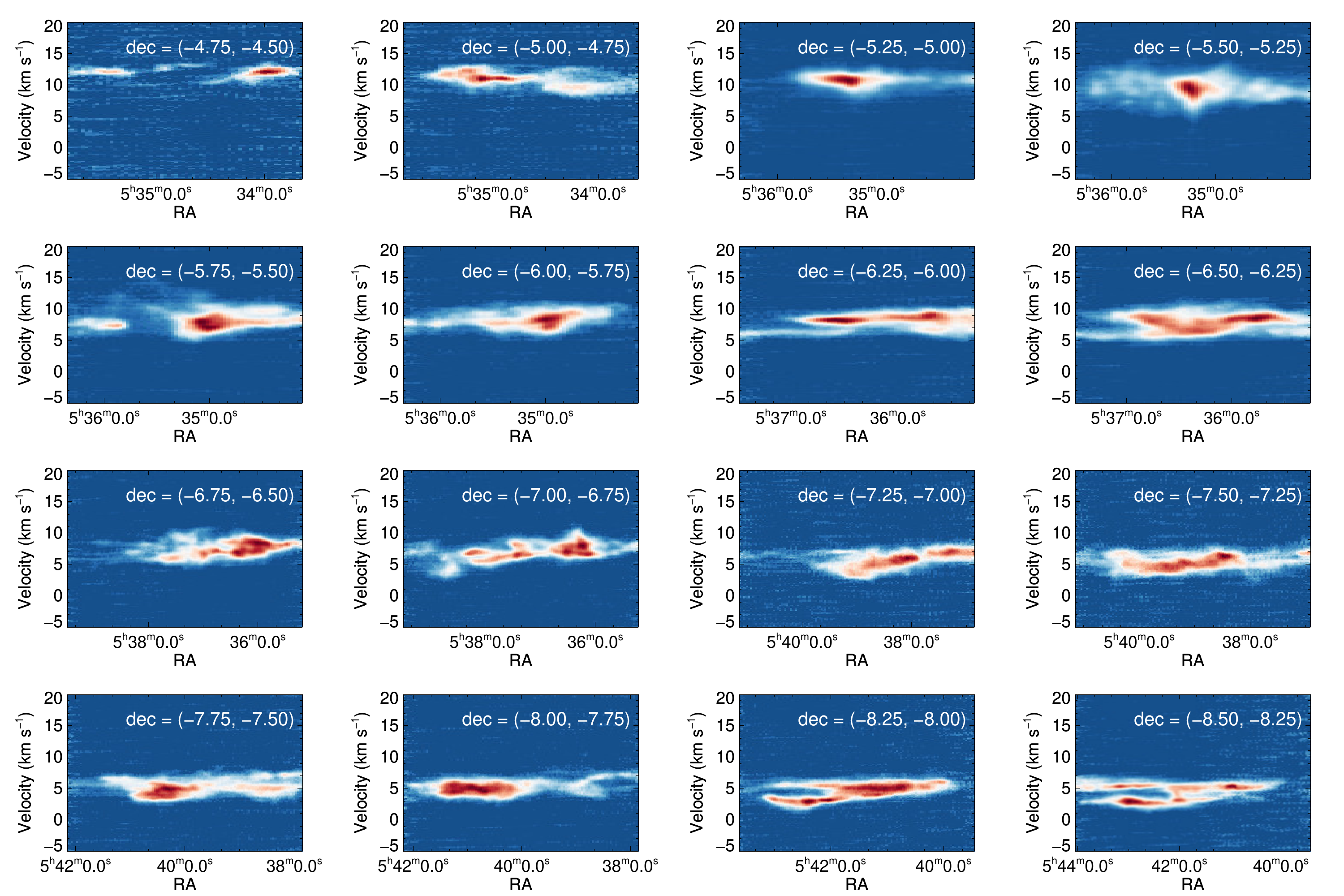}
	\caption{Same as Figure \ref{Fig6} but for the $^{13}$CO emission.}
	\label{FigB2}
\end{figure}

We present the velocity channel maps, p-v diagrams of the $^{13}$CO $J=1-0$ emission in Figs. \ref{FigB1} and \ref{FigB2}, respectively.

\clearpage
\bibliographystyle{raa}
\bibliography{Orion_RAA_arxiv.bbl}
\label{lastpage}
\end{document}